\documentclass[sigconf]{acmart} 
\AtBeginDocument{%
  }

\copyrightyear{2026}
\acmYear{2026}
\setcopyright{cc}
\setcctype{by}
\acmConference[CHI '26]{Proceedings of the 2026 CHI Conference on Human Factors in Computing Systems}{April 13--17, 2026}{Barcelona, Spain}
\acmBooktitle{Proceedings of the 2026 CHI Conference on Human Factors in Computing Systems (CHI '26), April 13--17, 2026, Barcelona, Spain}
\acmPrice{}
\acmDOI{10.1145/3772318.3790500}
\acmISBN{979-8-4007-2278-3/2026/04}

\begin{document}

\title{Botender: Supporting Communities in Collaboratively Designing AI Agents through Case-Based Provocations}

\author{Tzu-Sheng Kuo}
\email{tzushenk@cs.cmu.edu}
\orcid{0000-0002-1504-7640}
\affiliation{
  \institution{Carnegie Mellon University}
  \city{Pittsburgh}
  \state{PA}
  \country{USA}
}

\author{Sophia Liu}
\email{sophiawliu@berkeley.edu}
\orcid{0009-0008-7746-0749}
\affiliation{
  \institution{University of California, Berkeley}
  \city{Berkeley}
  \state{CA}
  \country{USA}
}

\author{Quan Ze Chen}
\email{cqz@cs.washington.edu}
\orcid{0000-0002-6500-8922}
\affiliation{
  \institution{AI \& Democracy Foundation}
  \city{Seattle}
  \state{WA}
  \country{USA}
}

\author{Joseph Seering}
\email{seering@kaist.ac.kr}
\orcid{0000-0001-7606-4711}
\affiliation{
  \institution{KAIST}
  \city{Daejeon}
  \country{Republic of Korea}
}

\author{Amy X. Zhang}
\email{axz@cs.uw.edu}
\orcid{0000-0001-9462-9835}
\affiliation{
  \institution{University of Washington}
  \city{Seattle}
  \state{WA}
  \country{USA}
}

\author{Haiyi Zhu}
\authornote{Co-senior authors contributed equally.}
\email{haiyiz@cs.cmu.edu}
\orcid{0000-0001-7271-9100}
\affiliation{
  \institution{Carnegie Mellon University}
  \city{Pittsburgh}
  \state{PA}
  \country{USA}
}

\author{Kenneth Holstein}
\authornotemark[1]
\email{kjholste@cs.cmu.edu}
\orcid{0000-0001-6730-922X}
\affiliation{
  \institution{Carnegie Mellon University}
  \city{Pittsburgh}
  \state{PA}
  \country{USA}
}

\renewcommand{\shortauthors}{Kuo et al.}

\begin{abstract}
AI agents, or bots, serve important roles in online communities. However, they are often designed by outsiders or a few tech-savvy members, leading to bots that may not align with the broader community's needs. How might communities collectively shape the behavior of community bots? We present Botender, a system that enables communities to collaboratively design LLM-powered bots without coding. With Botender, community members can directly propose, iterate on, and deploy custom bot behaviors tailored to community needs. Botender facilitates testing and iteration on bot behavior through \textit{case-based provocations}: interaction scenarios generated to spark user reflection and discussion around desirable bot behavior. A validation study found these provocations more useful than standard test cases for revealing improvement opportunities and surfacing disagreements. During a five-day deployment across six Discord servers, Botender supported communities in tailoring bot behavior to their specific needs, showcasing the usefulness of case-based provocations in facilitating collaborative bot design.
\end{abstract}

\begin{teaserfigure}
  \includegraphics[width=\textwidth]{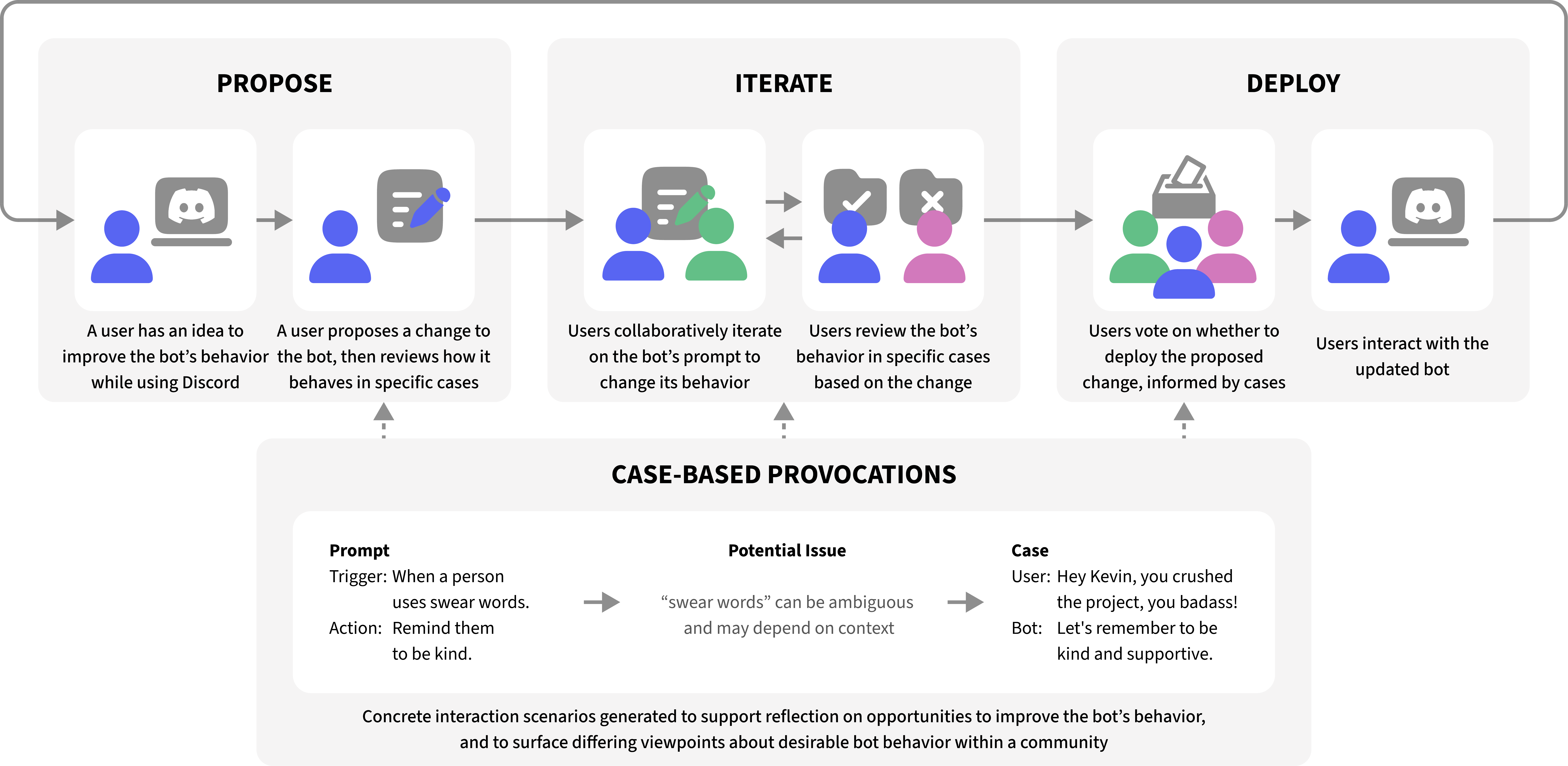}
  \caption{Botender is a system that supports users in collaboratively designing community bots. Using Botender, users can propose, iterate on, and deploy changes to a bot powered by LLM-based AI agents. Botender facilitates testing and iterating on the bot’s behavior through algorithmically generated case-based provocations: interaction scenarios designed to spark user reflection and discussion about desirable bot behavior. Users can iterate on the bot and make collective deployment decisions based on these cases.}
  \Description{An illustration of Botender’s user workflow features three main steps: propose, iterate, and deploy, with case-based provocation providing support throughout the entire process.}
  \label{fig:ch6:teaser}
\end{teaserfigure}


\maketitle

\section{Introduction}

Automated agents, often referred to as ``bots'' in online communities, play diverse and important roles across various community platforms \cite{seering2018social}. For example, conversational bots in Discord servers and Slack workspaces are commonly used to increase user engagement, such as by sending welcome messages \cite{hwang2024adopting,bali2023nooks}. Moderation bots \cite{doan2025design, seering2024chillbot}, like the Automoderator on Reddit \cite{jhaver2019human}, are widely adopted to enforce subreddit rules by sending warnings and removing posts that violate community standards. Utility-focused bots, such as ClueBot NG on Wikipedia \cite{geiger2010work, geiger2013levee}, help maintain the quality of community contributions by reverting damaging edits to articles \cite{kuo2024wikibench, halfaker2025collective, halfaker2020ores}. These bots not only perform specific task-oriented functions but also act as social actors in online communities \cite{seering2019beyond}, interacting with community members through conversations or other platform actions \cite{zhang2020policykit}. They are deeply integrated into the sociotechnical infrastructure of online communities and are vital for their growth, maintenance, and flourishing \cite{seering2018social, seering2019beyond, geiger2013levee}.

However, online communities often rely on third-party bots created by \textit{outsiders} who are not part of the community \cite{hwang2024adopting}, resulting in the adoption of bots that do not fully align with their specific needs and values \cite{hwang2024adopting}. 
This misalignment occurs because third-party bots typically offer limited customization options \cite{houde2025controlling}, and developers often lack the community-specific knowledge needed to adequately address tailored requests \cite{long2017could}. As outsiders who typically leave once a bot is developed, developers are also unable to assess the bots impact within communities and iteratively update the bot's design over time \cite{long2017could}. While some communities have a few members with the technical skills to build bots, the technical barrier to participating in bot design has created an unintentional hierarchy within the community and limits broader community participation in shaping the bots’ behavior \cite{geiger2014bots}. As a result, this can lead to undesirable, community-wide consequences that might have been avoided with greater community input \cite{halfaker2011don}.

Recent advances in Large Language Models (LLMs) have significantly reduced the technical barriers to bot design~\cite{houde2025controlling, petridis2024constitutionmaker, louie2024roleplay, liu2025proactive}. This presents an opportunity to support a more community-driven approach, where members are empowered to collectively shape the behavior of bots powered by LLM-based AI agents.\footnote{In this paper, we sometimes use ``AI agents'' and ``bots'' interchangeably. We generally use ``bot'' for the user-facing application, and ``AI agent'' when referring to the underlying implementation of a bot, which may involve one or more AI agents.
} By viewing bots as shared community infrastructure, such a community-driven approach could help ensure their alignment with a community's collective values and needs, rather than leaving this in the hands of outsiders or a small group of technical experts. However, supporting communities of \textit{non-AI experts} in \textit{collaboratively} designing LLM-based bots for themselves presents several open challenges. First, research shows that non-AI experts often focus narrowly on editing LLM prompts for a single interaction scenario \cite{subramonyam2025prototyping}, failing to test and account for unintended bot behavior across a range of relevant scenarios \cite{zamfirescu2023johnny}. Moreover, without a coordinated process, differing opinions among community members on how bots should behave can lead to difficulties in achieving consensus and effective collaboration \cite{fazelpour2025value, sorensen2024roadmap}. Finally, to support wider community participation, as emphasized in past studies \cite{kuo2024wikibench}, this design process has to be tightly integrated into existing community platforms. These challenges highlight a clear research gap in supporting the participatory design of AI agents within community contexts \cite{seering2019beyond, long2017could, hwang2024adopting}.

In this work, we present Botender, a system that enables online communities to collaboratively design and \textit{tend} to their bots over time. As illustrated in Figure \ref{fig:ch6:teaser}, Botender enables users to collaborate on (1) \textit{proposing} desired changes to the bot’s behavior, (2) \textit{iterating} on the design of prompt-based bot instructions to operationalize desired changes, through collaborative editing and testing, and (3) \textit{deploying} changes once users reach some level of consensus on its readiness. Botender supports this iterative, collaborative approach to prompt design through algorithmically generated \textit{case-based provocations}. These are concrete cases that illustrate how the bot would behave in concrete interaction scenarios, selected to \textit{probe the boundaries} of users' intents by provoking user reflection on (1) potential gaps between users' desired bot behavior and the behavior yielded by a given prompt, and (2) potential sources of disagreement about desirable bot behavior within a community. These provocations are designed to support iterative and collaborative prompt design by highlighting a given prompt's behavioral consequences in interaction scenarios that users may not initially consider. Finally, Botender is designed to be deeply integrated within community platforms to reduce barriers to participation and encourage broader community collaboration in bot design.

We first validated Botender's case-based provocation approach through an online experiment and then studied how communities use Botender in practice through a multi-day field study. The validation study focused on evaluating the effectiveness of Botender’s case-based provocation algorithm. The results suggest that, compared to a generic test case generation approach, participants found that Botender’s case-based provocations revealed more opportunities to improve the bot, and had greater potential to surface differing views about desired bot behavior among community members. The field study aimed to understand how communities use Botender by deploying the system in six real-world Discord communities over a period of five days. During this period, Botender supported participants in tailoring bot behavior to meet the unique needs and norms of their own communities, showcasing the usefulness of case-based provocations in facilitating collaborative bot design.

Overall, this work contributes Botender, a system that supports communities in collaboratively designing community bots through case-based provocations. Building on findings from an algorithm validation study and a five-day field deployment, we discuss opportunities for future HCI research to better support community-driven bot design.

\section{Related Work}\label{ch6:sec:related_work}

The study of bots in community contexts has been a major focus in HCI research \cite{seering2018social}. In this section, we first discuss the critical roles bots play in communities and the opportunity to support more collaborative, community-driven approaches to bot design. We then review existing work on supporting end users in designing bots, with an emphasis on recent efforts involving LLM-based agents. Finally, we discuss past work on the use of concrete cases to support iterative and collaborative design.

\subsection{The Roles of Bots in Online Communities}\label{ch6:sec:related_work:community}
Bots serve a variety of important roles across different community platforms \cite{seering2018social}. 
Bots in socially-focused communities, such as Discord servers, usually take on more \textit{socially-oriented} roles \cite{seering2020takes}, while in professional groups like Slack workspaces, they are typically designed to perform \textit{task-oriented} functions \cite{ashktorab2023sme}. Research literature often refers to bots that interact with multiple users simultaneously as \textit{multi-party} or \textit{polyadic} bots \cite{seering2019beyond, zheng2022ux}. This contrasts with \textit{dyadic} bots, which primarily engage in one-on-one interactions. Prior research has established taxonomies that categorize the types of content polyadic bots provide to user communities \cite{long2017could, seering2018social}. For example, Seering et al. classify bot content into five categories \cite{seering2018social}: sharing information\footnote{E.g., WikiBot on Discord links Wikipedia articles to messages: \url{https://www.wikibot.de}}, sending moderation warnings\footnote{E.g., Automoderator on Reddit sends moderation warnings to users \cite{jhaver2019human}}, facilitating user engagement\footnote{E.g., MEE6 on Discord welcomes newcomers to the server \cite{lee2025mapping}}, promoting community-approved advertisements\footnote{E.g., Nightbot on Twitch promotes advertisements for streamers’ merchandise \cite{seering2020takes}}, and running mini-games\footnote{E.g., Mudae on Discord facilitates anime character collection games \cite{hwang2024adopting}}. Communities typically search for existing bots that roughly meet their needs \cite{hwang2024adopting}, while HCI researchers have developed a variety of unique bots with specialized roles and functions to address gaps not covered by existing options \cite{seering2024chillbot, bali2023nooks, zhang2018making, savage2016botivist, kim2020bot, liu2025proactive}.

Although many bots are available on the market, communities often struggle to find the ``right'' bot for their needs and frequently end up adopting bots that misalign with their specific requirements \cite{geiger2010work,jhaver2019human, seering2019moderator}. This misalignment occurs because existing third-party bots typically offer limited customization options, and developers often lack the community-specific knowledge and capacity to address tailored requests over time \cite{hwang2024adopting}. For example, a study of more than two thousand requests on /r/requestabot, a subreddit that connects bot requesters with developers, found that external developers who lack community-specific knowledge often struggle to fully understand requests and are unable to develop bots that are tailored to the needs of individual communities \cite{long2017could}. While some communities do have members with both the technical skills to build bots and insider knowledge of the community’s needs, the technical barrier to participating in bot design significantly limits broader community participation in shaping the bots' behavior, which can result in undesirable, community-level consequences that might have been avoided with greater community input \cite{kraut2012building, geiger2014bots}. For example, bots created by technically inclined Wikipedia patrollers to revert potential vandalism have unintentionally discouraged new contributors, undermining efforts to retain them \cite{halfaker2011don}. Overall, these challenges emphasize the need for research into tools and processes that support more participatory approaches to bot design, allowing communities to collaboratively design bots that better meet their collective needs and values \cite{seering2019beyond, long2017could, hwang2024adopting}.

\subsection{Supporting End-Users in Prompt Design}\label{ch6:sec:related_work:prompt}
HCI research has a rich history of empowering end-users without technical skills to design AI systems tailored to their specific needs \cite{fails2003interactive}. Early work on interactive machine learning (iML) and machine teaching develop tools and processes for individual, non-technical users to design traditional ML models \cite{amershi2014power}, often for classification tasks \cite{chen2018anchorviz, lee2024clarify}. Some of this work has specifically focused on supporting the \textit{collaborative} design of ML models \cite{tseng2023collaborative, hong2020crowdsourcing}. These efforts have focused mainly on supporting end-users in collecting more diverse datasets to train more robust classification models.

Recent advances in LLMs have significantly lowered the technical barriers to designing AI systems capable of more complex tasks \cite{suresh2024participation}, such as AI chatbots. This presents an opportunity for communities to design their own LLM-based bots by writing prompts in natural language, without the need for coding. However, crafting effective prompts remains a challenging and unintuitive task for end-users \cite{zamfirescu2023johnny}. From writing prompts from scratch and iterating on them to assess their downstream impact, prior studies have identified several failure points in prompt design \cite{zamfirescu2023herding}. For example, a frequently encountered challenge is that non-AI experts often focus on iterating on their prompts for a specific interaction scenario they have in mind, without considering how these iterations might affect other scenarios \cite{subramonyam2025prototyping}. As a result, prompt iterations can unintentionally worsen outcomes for scenarios previously considered but not revisited, or for relevant scenarios that users had not even considered~\cite{zamfirescu2023johnny}. Insufficient consideration of how a bot might behave across diverse interaction scenarios can lead to prompts that are overly ambiguous, making it difficult for LLMs to distinguish between meaningfully different scenarios~\cite{chen2024your}. On the other end of the spectrum, this can also lead to prompts that are worded in a way that is overfit to a specific scenario, so that the LLM is only able to handle a narrow set of interaction scenarios~\cite{subramonyam2025prototyping}. Finally, this can lead to prompts that cause unintended downstream consequences in interaction scenarios a user had not considered~\cite{wang2024farsight,zamfirescu2023johnny}.

Prior work has explored a range of ways to support non-technical end-users in prompt design~\cite{arawjo2024chainforge, petridis2024constitutionmaker, louie2024roleplay, wu2022ai, ma2025should}, with some tools specifically created to address the aforementioned challenges. For example, to help address common pitfalls like writing overly ambiguous prompts, prior work creates a prompt coach that directly asks novice users high-level questions for the user to reflect on~\cite{chen2024your}, such as ``Is your prompt detailed enough?''. Other work provides users with direct recommendations on how they might edit their prompts to avoid potential undesirable social consequences~\cite{santana2025responsible, santana2025can}. Meanwhile, work such as Wordflow~\cite{wang2024wordflow}, PromptSource~\cite{bach2022promptsource}, and FlowGPT~\cite{li2024flowgpt} leverage the wisdom of the crowd by enabling individual users to upload their prompts to a shared repository and download prompts from others, helping users address scenarios they might not have considered on their own. However, since these methods are primarily intended for individuals to create personalized prompts rather than enabling groups to collaboratively develop prompts they use and rely on together, they do not facilitate direct collaboration on prompt design.

Enabling groups to collaboratively design prompts for shared LLMs or AI agents remains an underexplored area of research~\cite{houde2025controlling,huang2024collective}. Prior work such as PromptHive~\cite{reza2025prompthive} and CoPrompt~\cite{feng2024coprompt} has developed specialized interfaces that enable domain experts to collaboratively design prompts for their specific needs. For example, PromptHive allows mathematics educators to load homework problems, write prompts to generate homework hints, and share these prompts with colleagues via a shared library, where others can download, reuse, and refine them~\cite{reza2025prompthive}. CoPrompt enables programmers to share prompts with collaborators and request prompts directly within the programming IDE~\cite{feng2024coprompt}. In contrast to these approaches, which primarily focus on supporting prompt sharing, Botender is aimed at facilitating the collaborative design process itself through the use of concrete case-based provocations, as we will describe later. More closely related to our context, Koala is an LLM-based chatbot that participates in group discussions on Slack and allows participants to adjust four pre-defined settings via radio buttons to customize high-level aspects of the bot's behavior, such as its level of proactiveness (high, medium, or low)~\cite{houde2025controlling}. However, this work does not enable groups to collaboratively author prompts to specify desired bot behavior in detail. This limits communities' ability to customize bots to meet their specific needs.

Building on prior work, Botender aims to support communities in directly collaborating on iterative prompt design through the use of case-based provocations. These provocations are concrete interaction scenarios generated to support reflection on opportunities to improve the bot's behavior, and to surface differing viewpoints about desirable bot behavior within a community. Most closely related to this concept is a feature in the Gensors system, which automatically generates ``edge cases'' to help users stress test visual sensing models in situations they might not have considered, to help them to identify unanticipated failure modes~\cite{liu2025gensors}. In contrast to Gensors' edge cases, which help individual users debug visual sensors, Botender’s case-based provocations are designed to promote collective reflection on bot behavior across diverse social interaction scenarios---tailored to highlight concrete consequences of known pitfalls in novice prompt design. Botender uses these concrete cases as common ground to facilitate collaborative prompt design. For example, as discussed in Section~\ref{ch6:sec:botender}, by voting on cases users can discover where their expectations of bot behavior may differ. The use of cases as common ground in Botender is inspired by their documented success in supporting iterative, collaborative design in other contexts, as discussed in the next subsection.

\subsection{Using Cases to Support Reflection in Collaborative Design Processes}\label{ch6:sec:related_work:case}

Cases have served as a medium for design and deliberation across many fields \cite{aamodt1994case, epstein2014value, lehoux2020anticipatory, wright2020policy, kuo2023understanding}, including HCI research. Building on prior HCI scholarship, we understand cases as concrete scenarios within specific problem situations that make abstract concepts tangible for reasoning, communication, and negotiation \cite{chen2025case, kuo2025policycraft}. For example, to establish common ground for public policy deliberation, HCI research has used concrete cases to illustrate how individuals might be affected by such policies, which are typically more abstract and difficult to reason about compared to specific cases \cite{kuo2025policycraft}. HCI researchers have also brought strategically-selected legal cases that are more accessible to the general public before the judicial system to challenge existing laws and advocate for legal reform \cite{kirkham2023legal, yang2024future}. Similarly, in online communities such as Wikipedia and Reddit, reflecting on concrete cases of user content enables moderators to refine moderation rules to account for scenarios they had not initially considered when creating these rules \cite{park2022social, halfaker2025collective, kuo2024wikibench, fan2020digital, koshy2025venire, chen2023judgment}. Across these contexts---public policy, law, and moderation rules---concrete cases provide a valuable common ground for reflection and discussion, to inform policy iteration. This broader idea of iteratively developing general principles based on judgments about concrete cases aligns with the \textit{method of reflective equilibrium}, proposed by political philosopher John Rawls \cite{knight2017reflective}. In this method, the cases most useful for discussing and refining principles are often not those that conform to existing principles, but rather those where case-level judgments conflict with current principles, or cases that elicit divergent judgments among discussants~\cite{kuo2025policycraft}. Such cases push the group to reflect on their perspectives, and iteratively refine high-level principles and/or case-level judgments until they are in alignment (or equilibrium) \cite{chen2025case, chen2023judgment}.

In the context of LLM-based AI agent design, the principles that govern an agent’s behavior are commonly expressed as prompts. This parallel motivates us to explore how supporting collaborative prompt design can benefit from approaches in other domains, such as public policy, where concrete cases are used to facilitate deliberation and collaborative policy design. For example, PolicyCraft is a system that supports communities in collaboratively proposing, critiquing, and revising regulatory policies through discussion and voting on concrete cases \cite{kuo2025policycraft}. These cases present specific, hypothetical actions by community actors (e.g., community members, businesses, government entities) and allow others to vote on and discuss whether they believe such actions should be permitted in their community. The community then revises its regulatory policies through this collective discussion and consensus. In this work, we explore whether a similar case-based approach can facilitate the collaborative design of AI agent behavior, enabling community members to collectively reflect upon and discuss how they would want a community bot to behave. In contrast to PolicyCraft's focus on manually-written cases by community members, Botender explores the idea of automatically, dynamically generated \textit{case-based provocations} that support communities in identifying areas of disagreement and opportunities for bot improvement throughout an iterative, collaborative design process.

\section{Design Goals}
Based on our review of prior work, we synthesized the following four design goals for systems that aim to support the collaborative design of AI agents in community settings.
\vfill

\begin{itemize}

\item [\textbf{D1.}] \textbf{The system should facilitate a coordinated process for agent design that supports effective collaboration and enables meaningful collective action.} Even within a community that shares broad norms and values, individual members may hold differing views on how an ideal agent should behave and how to design it accordingly \cite{fazelpour2025value, chen2023judgment}. The system should facilitate identifying potential sources of disagreements, enable community members to discuss their ideas, and support collective decision-making through a process perceived as legitimate by the community \cite{shaw2014computer, kuo2025policycraft, salehi2015we}.
\vfill

\item [\textbf{D2.}] \textbf{The system should encourage users to assess the broader impact of their design ideas to support iterative prototyping.} 
As discussed in Section \ref{ch6:sec:related_work:prompt}, when designing AI agents, non-AI experts often focus narrowly on refining the agent's behavior in a \textit{single} interaction scenario, overlooking the broader impact of their prompt designs across a variety of interaction scenarios \cite{subramonyam2025prototyping}. As a result, prompt iterations can unintentionally worsen outcomes for other relevant scenarios that users had failed to consider \cite{zamfirescu2023johnny}. To address this, systems should support users in considering how their design may affect a wider range of cases they may not have initially anticipated, to inform design iteration.
\vfill

\item [\textbf{D3.}] \textbf{The system should support regression testing to help users prevent the reintroduction of previously resolved issues during iterative agent design.}
As discussed in Section \ref{ch6:sec:related_work:prompt}, non-AI experts often focus on refining an agent’s behavior in the interaction scenario they are currently considering, overlooking how their refinements may impact interaction scenarios they had previously considered \cite{subramonyam2025prototyping, zamfirescu2023johnny}. As a result, prompt iterations may unintentionally reintroduce undesirable agent behaviors in previously addressed scenarios. To mitigate this issue, prior HCI research \cite{zamfirescu2023johnny} suggests that systems should support regression testing---a practice from software engineering to ensure that new code does not break existing functionality \cite{elbaum2000prioritizing, rothermel1997safe, elbaum2014techniques}. In the context of collaborative AI agent design, this involves re-running the agent in previously resolved scenarios after making design changes to ensure it still behaves as the community intends.
\vfill

\item [\textbf{D4.}] \textbf{The system should be integrated into existing community platforms to promote broader community participation.} Designing an AI agent within a community context requires additional effort beyond community members' regular activities. To reduce participation barriers, a key strategy from past research is to directly integrate the system into the community platform \cite{kuo2024wikibench, irani2013turkopticon, halfaker2020ores}. This integration should provide multiple ways for members to contribute and let them decide how much effort they want to invest. It should also help direct their attention to the tasks that would most benefit from community input \cite{kuo2024wikibench}.
\vfill

\end{itemize}

\section{Botender}\label{ch6:sec:botender}
Based on these design goals, we developed Botender, a system that enables the collaborative design of bots in online communities. While Botender’s system architecture is designed to support the creation of diverse AI agents across different community platforms, in this paper, we present the first version of Botender, which supports the design of single-turn, LLM-based conversational bots powered by AI agents for Discord servers. In the following subsections, we first introduce Botender's overall system and agent architecture. Next, we walk through Botender’s user interaction workflow, as illustrated in Figure \ref{fig:ch6:teaser}. This workflow enables users to collaboratively design their bot by \textit{proposing}, \textit{iterating} on, and \textit{deploying} tasks for the bot to perform within their community platform. We then describe Botender’s \textit{case-based provocation algorithm}, which generates provocative test cases to encourage user reflection on bot design during the iteration process. Throughout the section, we connect specific features of Botender’s design to the design goals outlined in the previous section, denoted as D1 to D4. Finally, we conclude with implementation details.

\subsection{System and Agent Architectures}\label{ch6:sec:botender:system_architecture}
\begin{figure*}[t!]
  \centering
  \includegraphics[width=\linewidth]{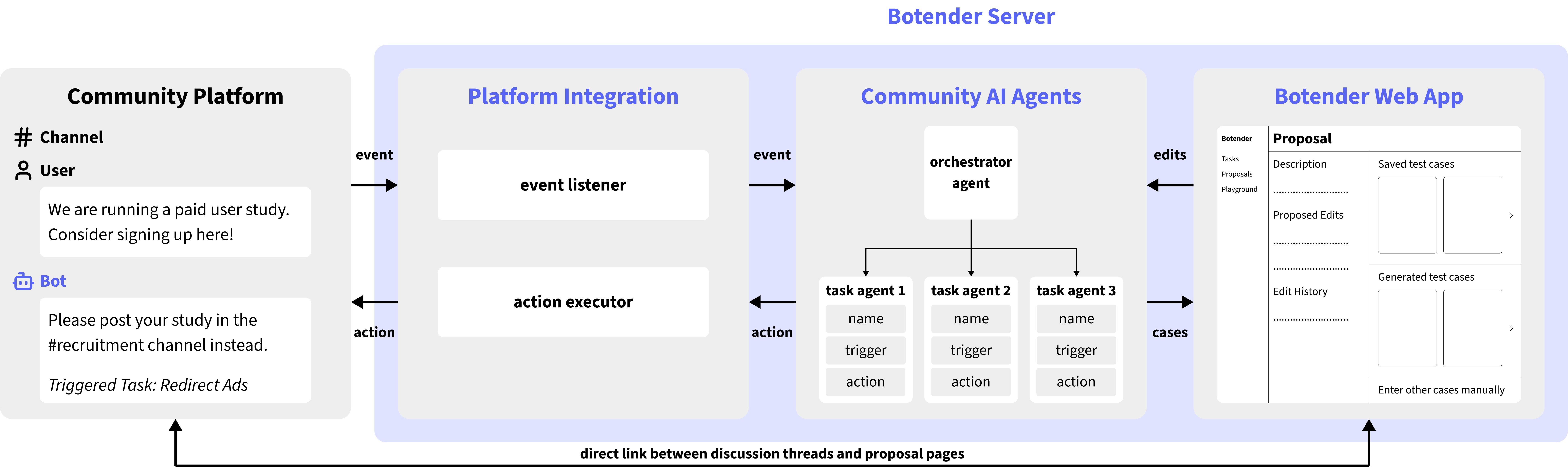}
  \caption{Botender's overall system and agent architecture. On the left, platform events are captured by Botender's always-running event listener, which translates them into information the agent architecture can understand. The orchestrator agent assesses each event and determines which, if any, task-specific agent is most relevant. The selected task-specific agent then generates an action instruction that is executed by Botender's platform action executor. On the right, Botender's website serves as the primary interface for users to collaboratively and iteratively design AI agents. This process generates concrete interaction scenarios that help guide further design iterations. The website is tightly integrated with the community platform, with each proposal directly linked to a dedicated discussion thread, as shown in Figure \ref{fig:ch6:discord}. This integration encourages broader community participation and discussion.}
  \Description{An illustration of Botender’s system architecture, showing four main components: community platform, platform integration, community AI agents, and Botender’s web app.}
  \label{fig:ch6:system_architecture}
\end{figure*}
Figure \ref{fig:ch6:system_architecture} depicts Botender’s overall system architecture and its integration with community platforms such as a Discord server. From the user's perspective, they install a single \textbf{bot} named Botender and interact with it on their community platform.\footnote{Like any Discord bot, only server admins can install Botender on their Discord server. After installation, admins can then grant other users permission to collaborate on designing the bot's behavior together. They can also change the bot's name from Botender to any name of their choosing.} The bot can perform a variety of \textbf{tasks}. The current version of Botender supports single-turn tasks: the bot handles one user message at a time, along with some meta information like the channel where the message was posted, and responds if any of its available tasks are triggered. For example, as shown on the left side of Figure \ref{fig:ch6:system_architecture}, if a user posts an advertisement in a certain channel, the bot recognizes this as relevant to its ``redirect ads'' task. It then replies to the user, suggesting they post the ad in a more appropriate channel. Users can work together to create any kind or number of tasks they want the bot to handle, tailoring its behavior to fit the unique needs and preferences of their community.

Behind the scenes, as shown on the right side of Figure \ref{fig:ch6:system_architecture}, the bot that users interact with is powered by multiple LLM-based agents, including an \textbf{orchestrator agent} and several \textbf{task-specific agents}, each corresponding to a task created by the users. This design aligns with best practices in agentic architectures \cite{fourney2024magentic, epperson2025interactive}, allowing each agent to focus on a single task and thereby enhancing overall transparency, controllability, and task outcomes \cite{wu2022ai, zhang2025chainbuddy}. Each task-specific agent includes a name, a trigger prompt, and an action prompt, which users can edit from Botender’s web interface. When the community platform induces an event detected by Botender's always-running listener, such as a user sending a message in a channel, the orchestrator agent assesses whether the event is relevant to any of the task-specific agents. This assessment is based on each agent's \textbf{trigger prompt}. If a specific task is deemed relevant, the corresponding agent will take the necessary action according to its \textbf{action prompt}, such as generating a response to the user's message. This action is then executed on the community platform.\footnote{Botender's system architecture is designed to support the bot's functionality beyond conversational interactions. For example, the event listener can monitor additional platform events, such as new users joining a server, and the system can execute a wider range of platform actions, such as banning users from the server. In the Discussion (Section \ref{ch6:sec:discussion}), we outline future directions for expanding the bot’s capabilities.} The complete system prompts for both the orchestrator agent and the task-specific agents are provided in Appendix \ref{ch6:sec:botender_appendix:agents}.

In the following subsections, we next walk through how users can use Botender to collaboratively design community bot behavior, by (1) proposing changes, (2) testing and iterating on prompts to operationalize proposed changes, and (3) deploying changes to the community platform (Figure~\ref{fig:ch6:teaser}).

\begin{figure*}[t!]
  \centering
  \includegraphics[width=\linewidth]{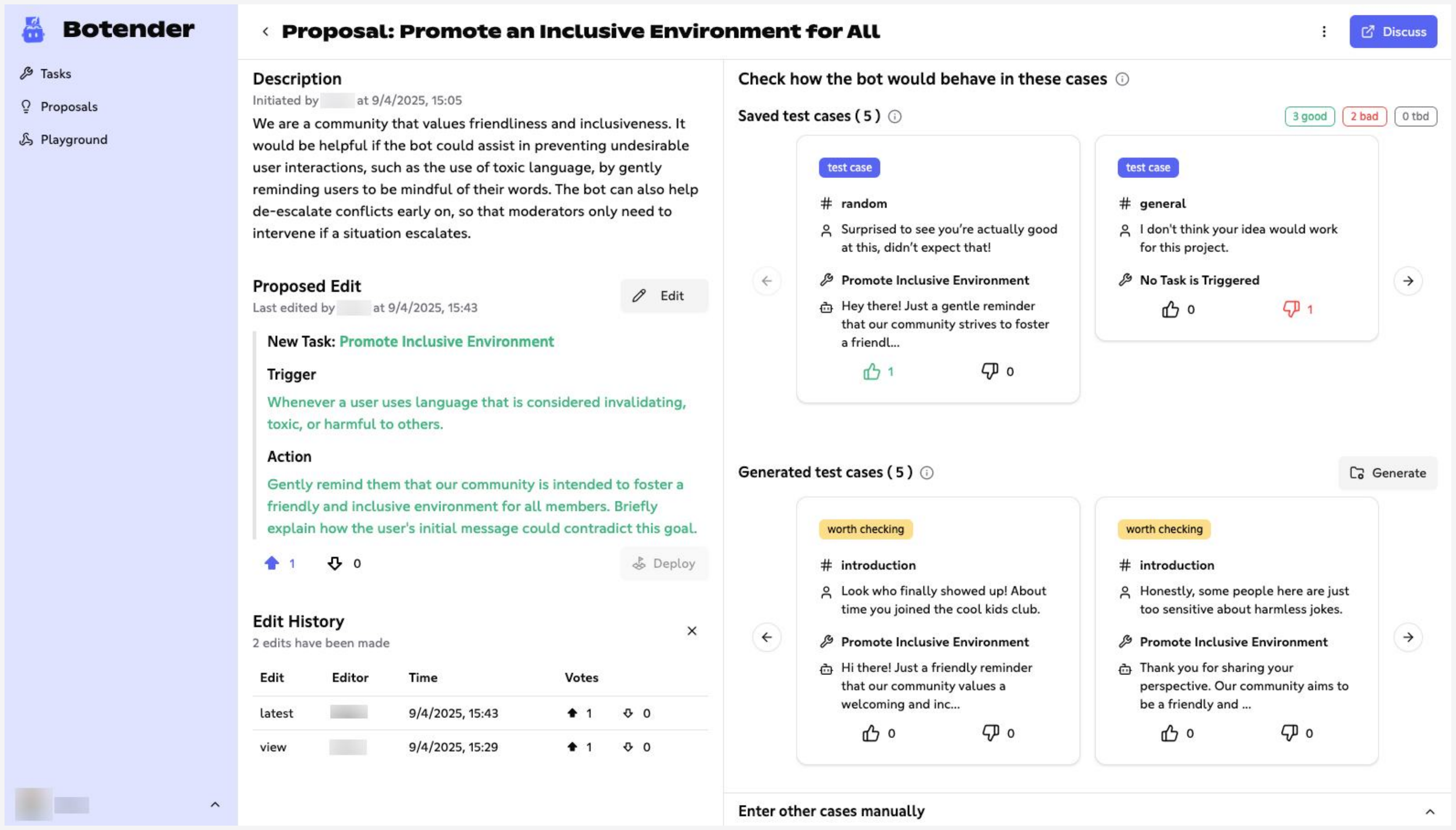}
  \caption{Botender's proposal page. The left navigation bar lets users switch between viewing all active tasks on their Discord servers, community proposals for desired changes, or experimenting with the bot in the playground without affecting their server. In the center, users see the proposal’s title, description, and the latest proposed edits to the bot’s tasks, such as adding a new task in this screenshot. Users can upvote or downvote to indicate their support for or opposition to deploying the latest edit. The bottom displays a full edit history, allowing users to compare edits with previous versions and the original task. On the right, test cases help guide collaborative decision-making. At the bottom are test cases automatically generated to provoke user reflection and discussion around the latest edit. Generated cases are saved if a user chooses to vote on the bot's response. At the top, members can review and vote on test cases that have previously been saved by community members. Clicking a test case opens a pop-up with case details, including how the bot’s responses for that case have changed across edits. Finally, users can click “enter other cases manually” to open a sheet where they can add custom test cases.}
  \Description{A screenshot of Botender’s proposal page.}
  \label{fig:ch6:proposal_page}
\end{figure*}

\subsection{Proposing Desired Changes to Bot Behavior}
To propose changes to the bot’s behavior, users can open Botender’s web interface directly from Discord.\footnote{It is common for more sophisticated Discord bots to have standalone websites for users to customize a bot's settings, rather than performing such customization within the Discord interface, so this is a familiar interaction for users.} Users may choose to initiate a \textbf{proposal} from the web interface, based on opportunities for bot improvement they notice during their everyday interactions on Discord. Alternatively, users may initiate a proposal based on observations they make while testing the bot's behavior using the \textit{playground} feature on Botender's web interface. Using the playground, users can freely test the behavior of the current version of the bot by simulating sending user messages in specific channels on their server. In the playground, users can also experiment with potential updates to bot behavior, and can then choose to submit a specific update as a proposal.

When creating a new proposal, users are asked to enter a title and a brief, high-level description of their desired changes. They are also encouraged to include a first attempt at operationalizing the change by either editing an existing \textbf{task} or creating a new one (see Figure~\ref{fig:ch6:proposal_page} for an example). As shown in Figure \ref{fig:ch6:edit_task}, each task consists of a pair of prompt fields: a trigger and an action. The \textbf{trigger} defines \textit{when} the bot should perform the task, while the \textbf{action} specifies \textit{what} the bot should do when the task is triggered. The task is also given a brief \textbf{task name} for future reference within the web interface and Discord, but this name does not affect the bot's behavior. As shown in Figure \ref{fig:ch6:discord}, the short name of the triggered task is displayed to users on Discord each time the bot replies. This allows users to better pinpoint and propose more precise edits to a specific task based on their observations of how the bot actually behaves within their community in accordance with that task.\footnote{These name, trigger, and action fields correspond to each task-specific agent's name, trigger prompt, and action prompt as discussed in Section \ref{ch6:sec:botender:system_architecture}.}

Once submitted, the proposal is created as a page visible to all community members, as shown in Figure \ref{fig:ch6:proposal_page}. If the original proposer has included a specific proposed edit with their proposal, they and any other visitors to the page are immediately shown a set of \textbf{test cases} illustrating how the bot would behave across a range of interaction scenarios (shown on the right side of Figure \ref{fig:ch6:proposal_page}), and are asked to review these test cases for unintended or undesirable bot behaviors. Each test case shows the channel name where a user message is sent, the user message itself, the name of the triggered task, and how the edited bot would reply to the user message. The test cases are divided into two sections: \textbf{generated test cases} and \textbf{saved test cases}. \textit{Generated test cases} are produced by Botender's case-based provocation algorithm (presented below in Section~\ref{ch6:sec:botender:algorithm}) with the aim of supporting user reflection on potential opportunities to improve the proposed edit (D2) and helping to surface disagreements among community members about whether the bot's response in a given scenario is appropriate (D1). The \textit{saved test cases} section allows the proposer or other community members to save test cases for later consideration and discussion. When saving a test case---whether a generated test case or a bot interaction from a user's manual testing---the user is asked to thumbs up or thumbs down the test case to indicate whether they think it is an example of a good or bad bot response. Other users can subsequently add their own votes to indicate their own perspectives. If a proposal is created based on a specific bot interaction the user observed in the playground, this case is automatically added to the saved test cases when the proposal is created.

When a community installs Botender on their Discord server, the system automatically creates a \textbf{\#botender} channel. This channel is primarily used for collaborating on bot design and is accessible only to admins by default, though permissions can be granted to other members if desired. When a new proposal is created, the system sends a \textbf{notification message} to the this Discord channel and creates a \textbf{discussion thread} linked to that message (D1, D4), as shown in Figure \ref{fig:ch6:discord}. Users can use this thread to discuss and coordinate their efforts, as well as follow the thread to receive notifications about proposal updates.

\subsection{Iterating on Bot Behavior}
\begin{figure}[t!]
  \centering
  \includegraphics[width=\linewidth]{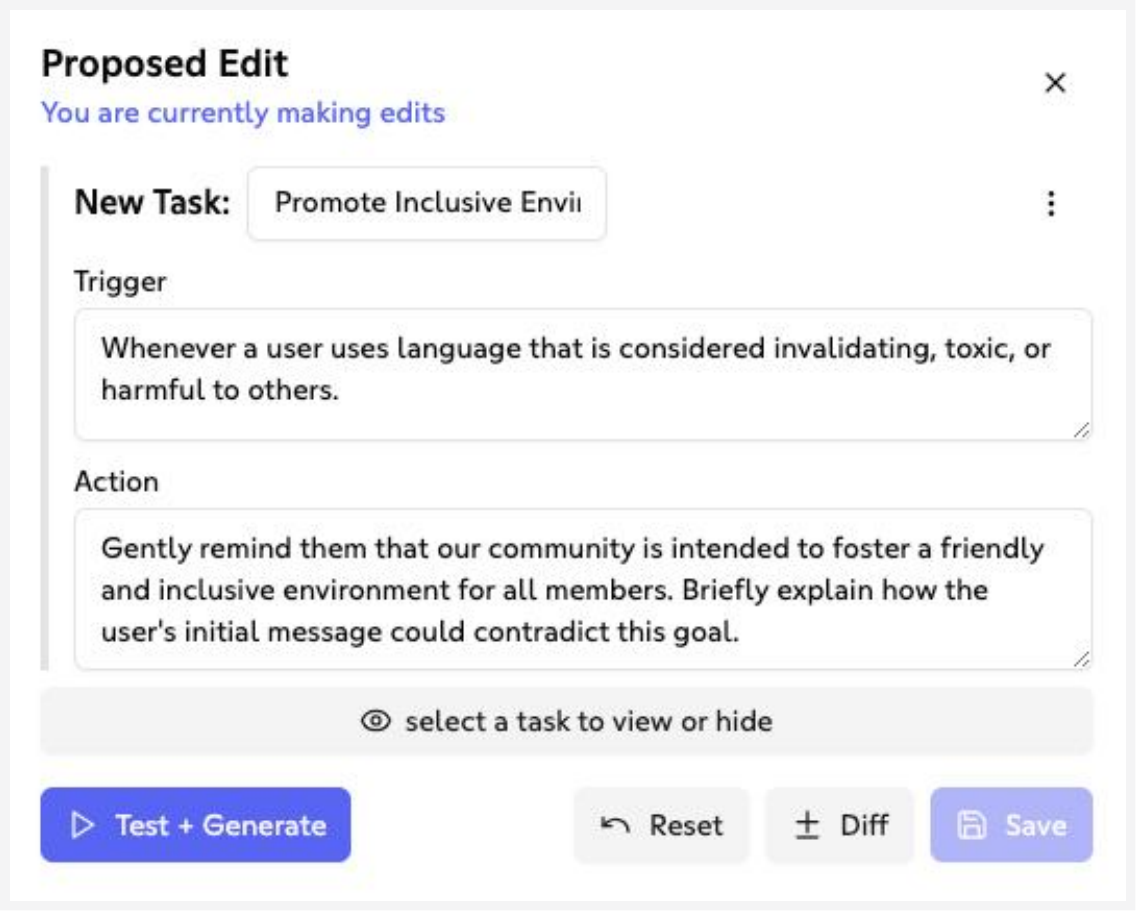} 
  \caption{After clicking the edit button on the proposal page, the original static text is replaced by this edit interface. Before saving edits, users are required to run ``Test + Generate'' to see how the bot would behave with their proposed edits.}
  \Description{An abridged screenshot of the edit interface on Botender’s proposal page.}
  \label{fig:ch6:edit_task}
\end{figure}

After a proposal has been created, community members can view the proposal page, including any saved test cases and newly generated case-based provocations for the latest version of the task. Users can review these test cases and add their votes on bot responses in saved test cases. The \textbf{counters} for ``good,'' ``bad,'' and ``tbd'' in the upper-right corner of a proposal page display the number of cases that have received a majority of thumbs up, thumbs down, or equal votes (Figure~\ref{fig:ch6:proposal_page}). They provide users with a quick overview of the community’s collective views on test cases. If users notice disagreements about desirable bot behavior that they wish to discuss, they can click the \textbf{``Discuss'' button} just above these vote counters, which brings users directly to the associated discussion thread for that proposal on Discord. 

Any user can make an \textbf{edit} to a task from a given proposal page, including creating new tasks or editing or removing existing ones. To do this, users click the ``Edit'' button next to the current proposed edit. To test their edits, users click the ``Test + Generate'' button, as shown in Figure \ref{fig:ch6:edit_task}. Botender then re-runs all saved test cases and displays the updated bot responses based on users' edits, to support regression testing (D3). Botender also generates new case-based provocations based on the user's edit (D2). Before a user is able to save their edit, they are asked to review the bot's behavior in the presented test cases, and must vote on \textit{at least one} of these test cases. This default threshold was chosen to promote user engagement with the test cases before saving and sharing proposed edits, while avoiding introducing too much friction into the collaborative editing process. However, the threshold may be customized to meet individual communities' needs. If the user notices a potential issue when reflecting on the test cases, they are encouraged to refine their edit and test again before saving. Once they are satisfied, they can save their edits to the proposal. 

\begin{figure*}[t!]
  \centering
  \includegraphics[width=\linewidth]{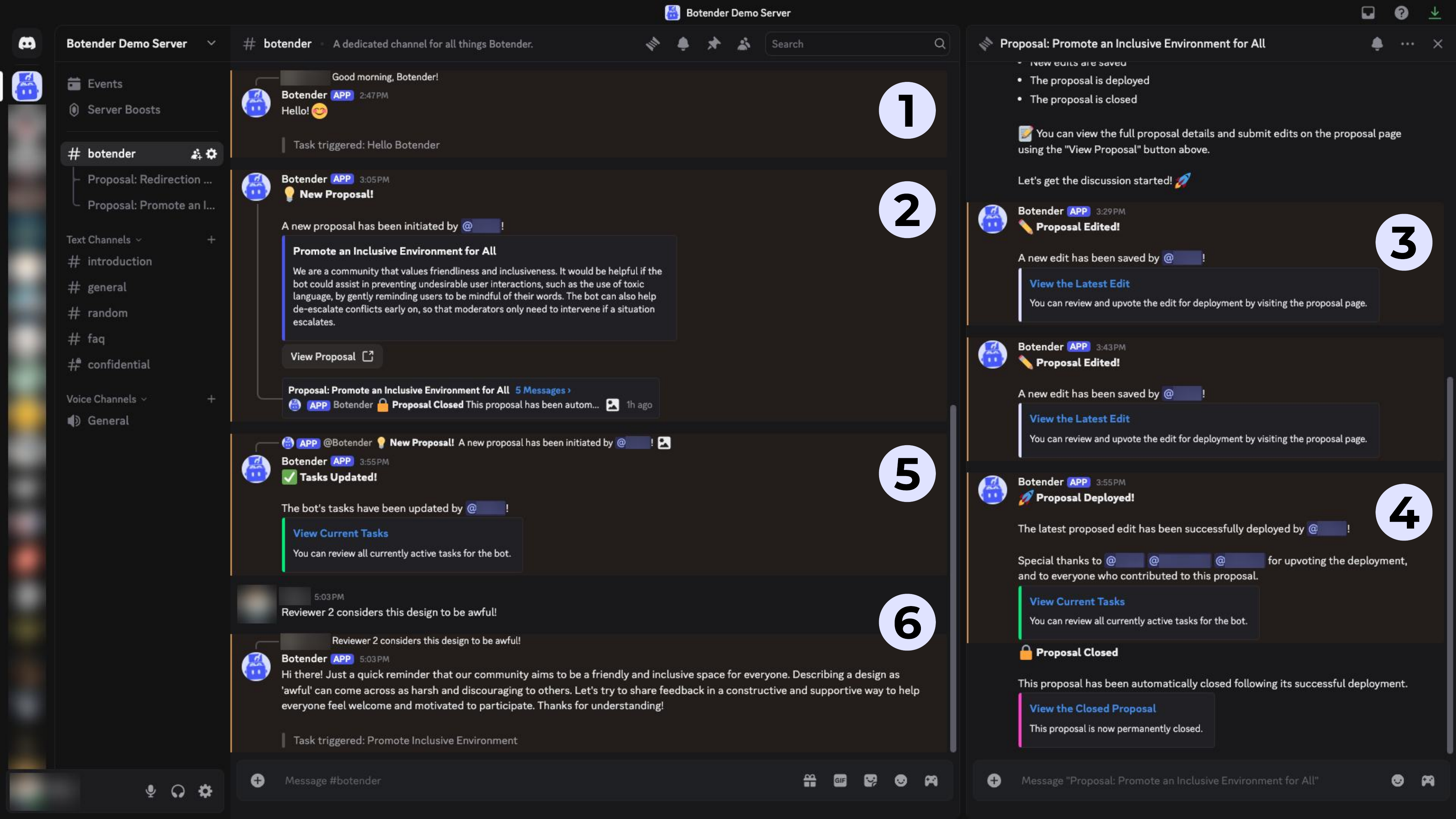}
  \caption{The Discord interface, highlighting Botender’s integration with the community platform. (1) By default, Botender replies "hello" to users who greet it in the \#botender channel, as defined by its default ``Hello Botender'' task. (2) When a new proposal is created, the system sends a notification to the \#botender channel, and (3) creates an associated discussion thread, as shown on the right, where users can discuss the proposal and receive notifications about saved edit updates. (4) Once a proposal is deployed, the system notifies the discussion thread, closes the proposal, and (5) and sends a message to the main \#botender channel. (6) The bot will then behave according to the latest deployed edit.}
  \Description{The Discord interface displaying Botender’s integration with the community platform, featuring six circular labels that correspond to the integration details described in the figure caption.}
  \label{fig:ch6:discord}
\end{figure*}

Once an edit is saved, the system sends a notification to the proposal’s discussion thread on Discord to encourage people to review the latest proposed edit (D1, D4), as shown in Figure \ref{fig:ch6:discord}. On the proposal page, they can then review the latest proposed edit and corresponding test cases. Users can also view the full \textbf{edit history} of a proposal, including edits to tasks and any additions or removals of saved test cases, and can revert changes as needed (D1). This design mirrors systems such as Wikipedia by making the edit history transparent \cite{dabbish2014transparency, stuart2012social, kuo2024wikibench}, which helps coordinate efforts and ensures accountability among editors.

\subsection{Deploying Updates to Bot Behavior}
At any point in this process, upon viewing the latest proposed edit and test cases for a given a proposal, users may choose to \textbf{vote} in favor of its deployment within their Discord server (D1). By default, each proposal requires at least three upvotes to be deployable, but this threshold can be adjusted to fit each community’s needs. Similar to saving edits, the system requires users to give at least one thumbs up or down on a saved test case before voting for deployment. As at other points throughout Botender's workflow---such as when creating or editing a proposed bot task---this encourages users to review how the bot would actually behave based on the proposed edit before casting their vote, rather than merely reading the edit.

Once a proposed edit reaches the deployment threshold, users can click the \textbf{``Deploy'' button} on the proposal page to deploy the proposed changes to the live bot on their Discord server. The system sends notifications to both the proposal discussion thread and the main \#botender channel to inform everyone that the bot’s tasks have been updated (D1, D4). The proposal is then closed.\footnote{If users choose not to deploy a proposal, they can close it but still have the option to reopen and edit it later if needed.}

\subsection{Case-Based Provocation Algorithm}\label{ch6:sec:botender:algorithm}
A key aspect of Botender is its support for collaborative iteration on bot design through the use of concrete test cases. Drawing inspiration from the use of cases as a medium for collaborative design and deliberation in various fields (Section \ref{ch6:sec:related_work:case}), Botender’s case-based provocation algorithm is designed to generate concrete interaction scenarios that spark collective reflection and discussion about desirable bot behavior, rather than simply \textit{validating} expected outcomes. The goal of these case-based provocations is to test the boundaries of users' intents in bot design (as expressed through LLM prompts) by (1) highlighting potentially undesired bot behavior in interaction scenarios users may not have considered; and (2) highlighting areas where there is potential for disagreement among different users. Based on past research, discussed in Section \ref{ch6:sec:related_work:case}, we expect such boundary-challenging provocations to be more useful in supporting prompt iteration than test cases aimed at simply validating expected behavior in expected scenarios.

Botender’s case-based provocation algorithm is designed to generate three broad types of cases, aimed at surfacing issues commonly overlooked by non-AI experts when designing LLM prompts, which can also be sites of disagreement among community members:
\begin{itemize}
    \item \textbf{Cases that highlight ambiguities in a prompt}: Non-AI experts often write LLM prompts containing ambiguous, underspecified phrases \cite{zamfirescu2023johnny}. Such phrases can be interpreted differently by both the LLM and by different people. For example, the phrase ``inappropriate language'' in a prompt leaves substantial room for differing interpretations about what exactly constitutes inappropriate language. If a user's interpretation of the prompt differs from that of the LLM, the agent may behave in ways that are unexpected to the user. Similarly, different users may agree on a prompt’s wording, only to realize that they have different expectations for how the agent should behave, upon seeing actual examples of bot behavior in different interaction scenarios. Concrete cases can help reveal such disagreements and encourage iteration on prompts to make them clearer and more specific.
    \item \textbf{Cases that highlight potential overly narrow wording in a prompt}: Non-AI experts also often write LLM prompts that focus too narrowly on defining an agent’s behavior for a single scenario, overlooking the broader impact of their prompt across a wider range of possible interaction scenarios \cite{zamfirescu2023johnny,subramonyam2025prototyping}. For example, a prompt that instructs the agent to identify a specific list of banned words (as is common in traditional moderation tools) may overlook how those words could be used legitimately in some contexts. Such a prompt may also fail to address the broader goal behind banning these words by missing related words or phrases that are not included in the list. Concrete cases covering a diverse range of relevant scenarios can encourage users to reflect and discuss how to iterate on an overly narrow prompt to better achieve their broader goals.
    \item \textbf{Cases that reveal potential unintended community-level consequences of a prompt}: Finally, as documented in prior HCI research, bots deployed in community contexts can sometimes lead to unintended community-level consequences, despite good intentions \cite{kraut2012building}. For example, community bots may inadvertently discourage participation by enforcing norms too strictly or crowding out opportunities for meaningful user contribution \cite{halfaker2011don}. Concrete cases can help users foresee potential unintended downstream consequences and iterate on the prompt before deployment to prevent them.
\end{itemize}

\begin{figure*}[t!]
  \centering
  \includegraphics[width=\linewidth]{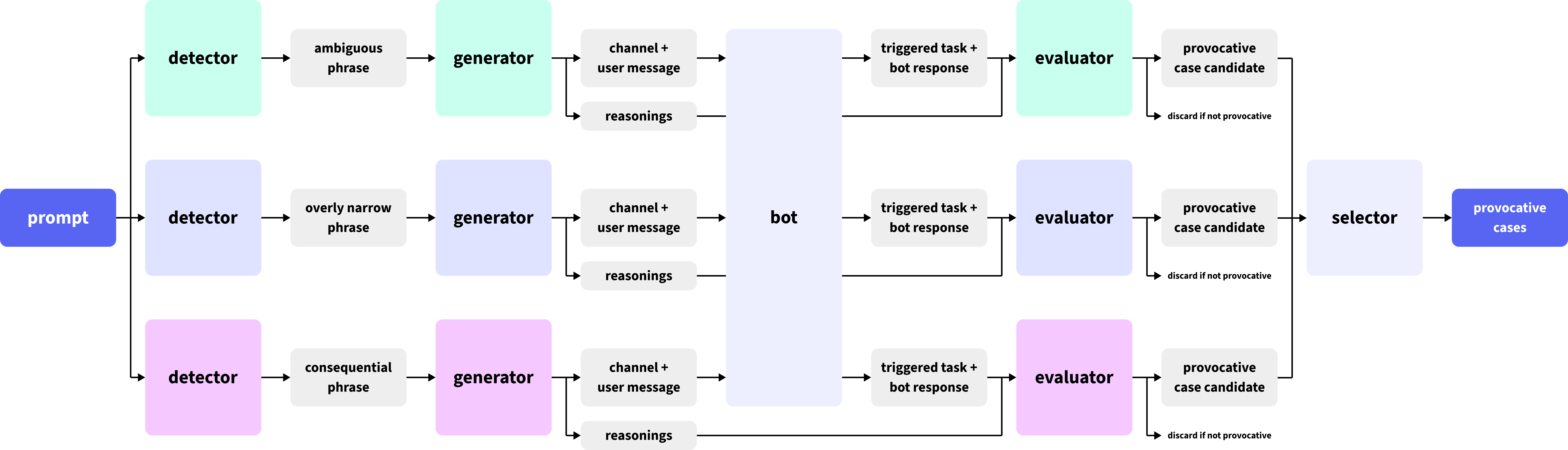}
  \caption{Botender’s case-based provocation algorithm uses three parallel LLM pipelines to generate provocative test cases that encourage user reflection and discussion on common prompt design pitfalls, including ambiguous language, overly narrow phrasing, or unintended downstream consequences for the community. Each pipeline includes its own detector, generator, and evaluator to generate relevant cases. The detector identifies phrases related to the targeted pitfall; the generator creates cases based on these identified phrases; and the evaluator checks how well the generated cases illustrate the pitfall. Finally, a selector chooses the most provocative cases from all case candidates. The prompts for all ten LLM modules, including each pipeline's detector, generator, and evaluator, as well as the final selector, are provided in Appendix \ref{ch6:sec:botender_appendix:algorithm}.}
  \Description{A flowchart illustrating the LLM modules that power Botender’s case-based provocation algorithm.}
  \label{fig:ch6:pipelines}
\end{figure*}

To generate these three types of cases to support user reflection and discussion, Botender's algorithm uses three separate LLM pipelines, one for each case type, as shown in Figure \ref{fig:ch6:pipelines}. Each pipeline consists of three modules: a detector, a generator, and an evaluator. The \textbf{detector} identifies specific phrases in the prompt, whether in the trigger or action prompt of a task, that may be too ambiguous, overly narrow, or could lead to unintended consequences. Based on the detected issue, the \textbf{generator} creates a channel name and user message that aim to concretely illustrate the problem, along with a \textbf{reasoning} of what the case aims to highlight. However, at this stage the generator does not yet know how a bot's will respond to the message. Therefore, after generating the channel and user message, these are sent to the bot (specifically, to the orchestrator agent and task-specific agents as described in Section \ref{ch6:sec:botender:system_architecture}) to obtain a response. The \textbf{evaluator} then reviews the complete case, including the actual bot response, to determine whether it effectively demonstrates the identified issue as described by the generator's reasoning. Only cases that pass the evaluator are output by each pipeline. Finally, all cases from the different pipelines are merged into a single pool, from which a \textbf{selector} module chooses a set of cases that are likely to be most useful (five in the current implementation) in promoting user reflection and discussion. Full details of the system prompts for individual LLM modules within the pipeline are provided in Appendix \ref{ch6:sec:botender_appendix:algorithm}.

\subsection{Implementation Details}
Botender is a full-stack, end-to-end system that individual communities on Discord can set up via an installation link, as is standard for Discord bots. Botender’s website has a fixed URL where Discord users can log in with their Discord accounts, but they can only view and design bots for servers where they are members and where Botender is installed. Botender is built with SvelteKit and hosted on Vercel, using shadcn-svelte components styled with Tailwind CSS on the frontend and Firestore databases on the backend. Since Vercel is a serverless platform, Botender’s always-on listener for Discord platform events is hosted separately on Railway. All the AI agents and LLM modules are powered by OpenAI’s GPT-4.1. The entire Botender codebase is open source and publicly available on GitHub.\footnote{\url{https://github.com/tskuo/Botender}}

\section{Algorithm Validation Study}\label{ch6:sec:validation_study}
We first conducted a validation study to understand how well Botender's \textit{case‑based provocation algorithm} generates test cases that support user reflection on desired bot behaviors. In Botender, these cases are aimed at (1) helping users identify opportunities to improve the bot's prompt and (2) helping to surface potential disagreements among community members about how the bot should behave. This validation study focuses on assessing how effectively Botender's case-based provocation algorithm achieves these goals.

\subsection{Recruitment}
We conducted the validation study through an online survey. In total, we recruited 90 participants on Prolific with experience in online groups or communities (e.g., Discord servers or Slack workspaces) where Botender is intended to be deployed. Each participant was compensated \$5 USD, and the median survey completion time was 13 minutes.

\subsection{Study Procedures}
Each participant was randomly assigned to review one of nine pre-selected bot prompts, which we referred to as bot ``instructions'' in the survey to avoid technical jargon. We selected these prompts to cover a variety of potential pitfalls in prompt design. Each prompt included both a trigger and an action, consistent with how tasks are specified in Botender. Additional details about the prompts used in the validation study are available in Appendix \ref{ch6:sec:validation_study_appendix:prompts}. 

Each participant reviewed a prompt and evaluated two sets of cases generated based on the following two conditions:
\begin{itemize}
    \item \textbf{Botender: Case-Based Provocations}: For each prompt, participants evaluated five cases generated by Botender’s case-based provocation algorithm, as described in Section~\ref{ch6:sec:botender}.
    \item \textbf{Baseline: Standard Test Cases}: For each prompt, participants also evaluated five algorithmically generated test cases relevant to the prompt but not specifically targeted to provoke critical reflection. By comparing against this baseline, we aimed to understand whether and how Botender's case-based provocations provide value beyond the \textit{general} benefits of encouraging people to reflect on concrete cases. See Appendix \ref{ch6:sec:validation_study_appendix:baseline} for the algorithm used to generate baseline cases.
\end{itemize}
Participants evaluated two sets of cases in random order. For each case, they rated how strongly they agreed or disagreed with the following two statements reflecting the goal of Botender’s case-based provocation algorithm:
\begin{itemize}
    \item \textbf{Controversialness}: I think \textbf{people may have differing opinions} on whether the bot's response in this case is appropriate.
    \item \textbf{Provocativeness}: I think this case \textbf{reveals opportunities to improve the instructions} that were given to the bot.
\end{itemize}
Participants also provided set-level ratings indicating their agreement with the following two statements and briefly described potential problems they saw with the bot's prompt, based on the cases in each set:
\begin{itemize}
    \item \textbf{Coverage}: This set of cases covers a comprehensive range of problems with the instruction.
    \item \textbf{Diversity}: This set of cases covers a diverse range of problems with the instruction.
\end{itemize}
Finally, participants selected the set of cases that revealed more potential problems with the prompt and thus better highlighted ways to improve the bot. They also briefly justified their choice. With a total of 90 participants, this sample size yielded 10 independent reviews per prompt and its corresponding cases. All prompts and cases that participants reviewed can be found in the supplementary materials.

\subsection{Study Results}
\begin{figure}[t!]
  \centering
  \includegraphics[width=\linewidth]{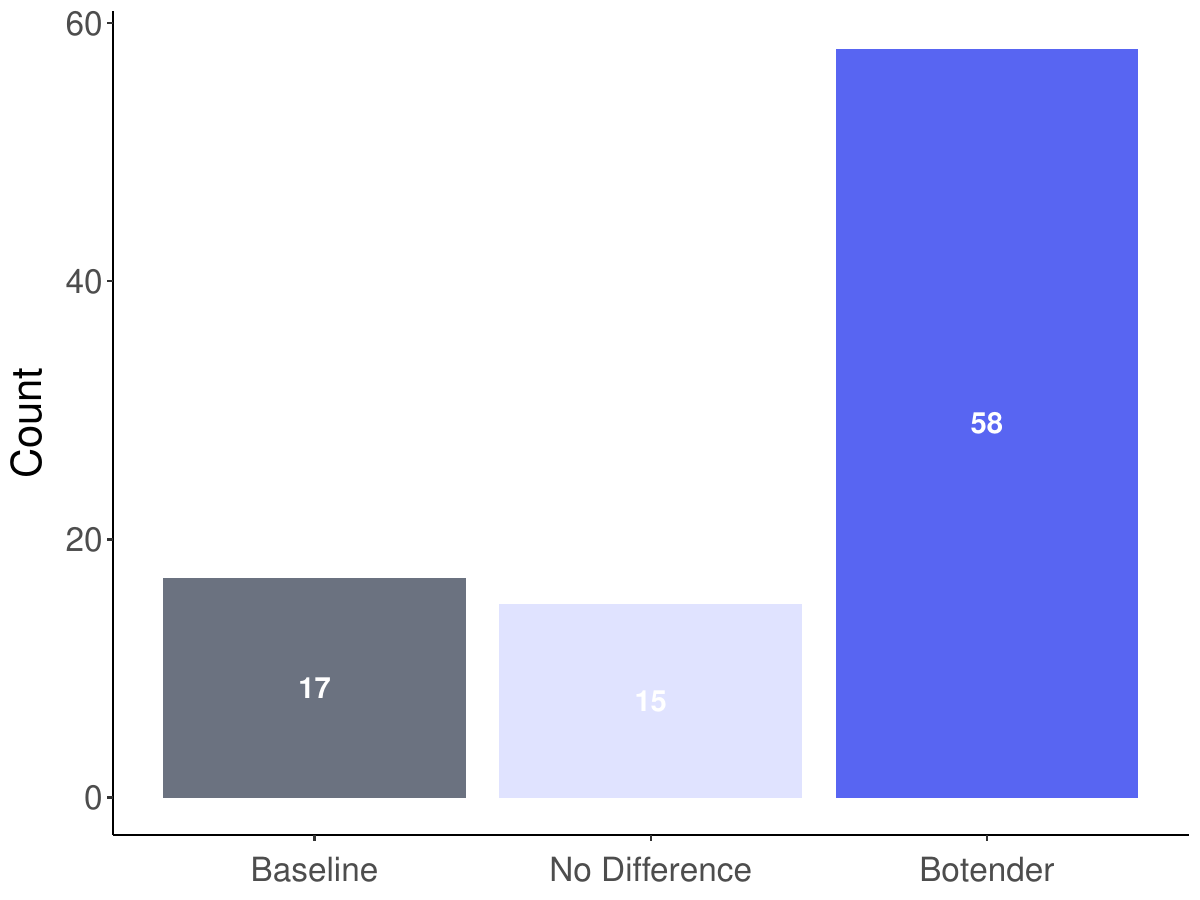} 
  \caption{The number of participants who reported that the sets of cases generated by the baseline vs. Botender's case-based provocation algorithm were more provocative (i.e., revealed more opportunities to improve the bot's instructions).}
  \Description{A bar chart comparing the number of participants who found the sets of cases generated by the baseline and by Botender’s case-based provocation algorithm to be more provocative. The results show 17 participants favored the baseline, 15 reported no difference, and 58 preferred Botender.}
  \label{fig:ch6:p_compare_all}
\end{figure}

\textbf{Overall, participants found that Botender’s case-based provocations revealed more opportunities to improve the bot's instructions.}
As shown in Figure \ref{fig:ch6:p_compare_all}, most participants selected the set of cases generated by Botender’s algorithm when asked which set revealed more opportunities to improve the bot's instructions, in a blind comparison. Participants’ justification of their choices\footnote{We denote participant IDs starting with V and the prompt a participant reviewed starting with P---for example, ``(V79, P1)''.} offered insight into the differences they perceived between the two sets of cases. For example, several participants noted that Botender's case-based provocations \textit{``surface a wider range of edge cases that the current instruction doesn’t handle well''} (V80, P8). Participants mentioned that these cases \textit{``are more vague and difficult for the bot to interpret''} (V79, P1), \textit{``show tricky situations the bot's instructions don't cover''} (V59, P7), and \textit{``highlight unclear criteria and over-triggering, thus better exposing instruction weaknesses''} (V62, P2). Meanwhile, they found the standard test cases to be \textit{``more cut and dry''} (V4, P7), offering either \textit{``obvious examples of red flags to the bot''} (V19, P1) or cases that \textit{``shows the bot doing its job correctly''} (V72, P2).

After reviewing Botender's case-based provocations, participants identified a range of opportunities to improve the bot's instructions, including concerns about potential unintended community-level consequences of the bot’s responses, such as making users \textit{``feel unimportant, unheard, and excluded''} (V51, P4), or noted situations where \textit{``the bot is too direct in question[ing] the user and it comes across as arrogant''} (V23, P7). By contrast, after reviewing cases generated by the baseline, participants found that the set \textit{``covers cases where the bot should respond, which it does''} (V45, P5), or that the bot \textit{``answered but just needed to shorten them or make them more to the point''} (V27, P4). Some participants observed that \textit{``[Botender's] set had many more areas for improvement''} (V31, P5), or found the baseline \textit{``showed no noticeable errors or deviations, and it responded as everyone would expect''} (V87, P3). Overall, participants found that Botender's case-based provocations \textit{``[hit] deeper problems''} (V90, P7).

\begin{figure*}[t!]
  \centering
  \includegraphics[width=0.99\linewidth]{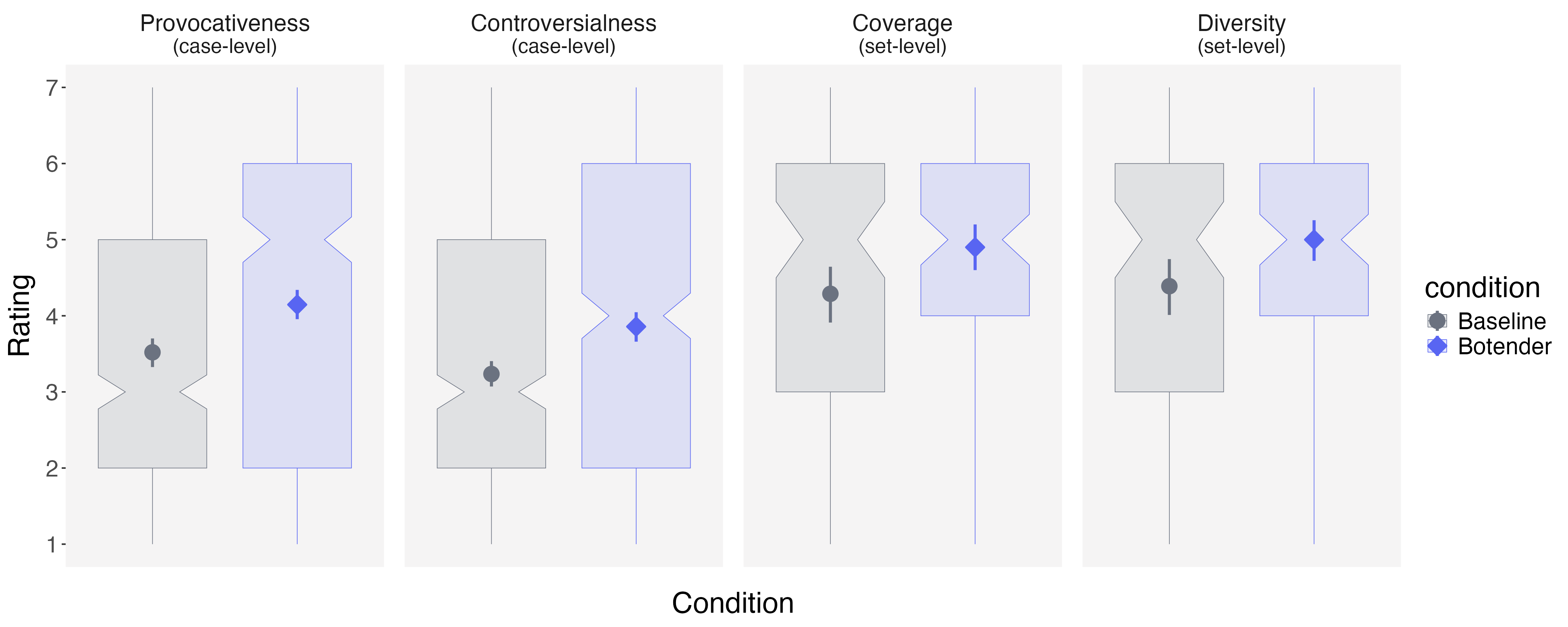}
  \caption{Participants’ ratings on various aspects of cases generated by Botender’s case-based provocation algorithm and the baseline algorithm. The figure shows notched box plots, with the notches indicating the medians, and the means and 95\% confidence intervals overlaid on the plots. The results show significant differences for all rated aspects.}
  \Description{A point range chart displaying the mean and 95\% confidence intervals for participants’ ratings on four aspects of the cases (provocativeness, controversialness, coverage, and diversity) generated by Botender’s case-based provocation algorithm compared to the baseline algorithm.}
  \label{fig:ch6:p_rating_all}
\end{figure*}

The finer-grained ratings participants provided at the case and set levels align with these interpretations. As shown in Figure \ref{fig:ch6:p_rating_all}, at the case level, participants gave significantly higher ratings ($p<0.001$) for both \textit{provocativeness} (the extent to which the case reveals opportunities to improve the bot’s instructions) and \textit{controversialness} (the extent to which they thought people might hold differing opinions about the appropriateness of bot behavior in a given case). At the set level, participants also gave significantly higher ratings ($p<0.01$) for both the coverage and diversity of Botender's case-based provocations. The effect size is significant but small at the individual case level—0.6 on provocativeness (3.5 for baseline and 4.1 for Botender) and 0.7 on controversialness (3.2 for baseline and 3.9 for Botender), likely due to variation across individual cases. However, at the aggregate set level, people overwhelmingly chose the Botender condition. Taken together, participants' judgments of provocativeness at the set level (Figure \ref{fig:ch6:p_compare_all}) versus at the individual case level (Figure \ref{fig:ch6:p_rating_all}) may indicate that participants found the case-based provocations most provocative when presented as a set, rather than in isolation. 

Overall, these results validate that Botender's case-based provocation algorithm generates sets of test cases that can better support user reflection on opportunities to improve the bot's instructions. This is further supported through findings from our field study, reported in the next section, where participants made heavy use of case-based provocations to collaboratively iterate on their prompts.

\section{Field Study}\label{ch6:sec:field_study}
To understand how people use Botender as an end-to-end system for collaborative AI agent design, we conducted a 5-day field study in real-world Discord communities. On Discord, each server is an online community where individuals with shared interests connect and interact. These interests range from casual hobbies like gaming and anime to professional topics such as programming or specialized spaces for customer service \cite{aquino2025discord}. Our goal in this field study was to understand how diverse Discord communities use Botender to collaboratively design and customize bots for their servers.

\subsection{Recruitment}
\begin{table*}
  \caption{Field study participant group demographics, including the number of participants in each group, the total number of community members in each group’s server where Botender was deployed, and a brief description of each server community.}
  \label{table:ch6:participants}
  \begin{tabular}{cccl}
    \toprule
    Group ID & Group Size & Server Size & Server Type\\
    \midrule
    G1 & 5 & 5 & A close-knit friend group server for hanging out and having fun \\
    G2 & 3 & 17 & A fan community for an indie music band to connect with their superfans \\
    G3 & 6 & 27 & A research lab led by a professor, with members including students and collaborators \\
    G4 & 6 & 50 & An offshoot of a larger community, created for members who share common interests in gaming \\
    G5 & 5 & 66 & A friend group and their close friends, primarily used for socializing and gaming\\
    G6 & 6 & 429 & A student organization within a university that organizes hackathons\\
    \bottomrule
  \end{tabular}
\end{table*}
We recruited six participant groups via social media, paper flyers, and word of mouth. These groups represented a diverse array of Discord communities, ranging from close-knit friend groups and student organizations to the fan community of an indie band. Table \ref{table:ch6:participants} provides details about each participant group.\footnote{During the study, other members of each server who were not participants did not use Botender to design bots.} All participants were active community members or admins in their servers, making them ideal candidates to design their own community bots using Botender. Each participant received \$100 USD for the field study, consistent with compensation provided in previous week-long HCI research (cf. \cite{zhang2018making, kuo2024wikibench}).

\subsection{Study Procedures}
The field study lasted five days for each group of participants. The study began with a synchronous onboarding session, where participants received guidance on how to use Botender and had the chance to try the system in a dedicated onboarding server. We also recorded a comprehensive system walkthrough for a few participants who were unable to attend the group onboarding session. After onboarding, participants installed Botender on their own Discord server and selected the start date for the five-day study period. We set two minimum participation requirements for the study. First, each participant needed to create or edit at least one proposal per day. Secondly, as a group, they were required to deploy at least three tasks tailored to their community’s specific needs, norms, and values by the end of the study. We kept the participation requirements minimal to provide them with the flexibility to decide when and how much they want to engage in bot design, while also ensuring that participants would have ample opportunities for interaction across the field study period (cf. \cite{zhang2018making, kuo2025policycraft}). We encouraged participants to design tasks that reflected their community’s unique norms and culture, rather than purely logistical tasks.

Each time participants edited a proposal on the proposal page, they answered a brief multiple-choice question about their motivation for making the edit. By analyzing the frequency of each selected option, we gained insights into what drives proposal editing. Participants could select one or more of the following six options to indicate that they were making an edit to address:
\begin{itemize}
\item [1.] specific saved test cases they saw
\item [2.] specific generated test cases they saw
\item [3.] specific cases they entered themselves manually
\item [4.] general issues that someone else raised
\item [5.] general issues they thought of themselves
\item [6.] other
\end{itemize}

At the end of the field study, all participants completed a post-study survey that included the following five statements. Participants rated their level of agreement with each statement on a scale from 1 to 7, where 1 indicates ``strongly disagree'' and 7 indicates ``strongly agree.'' They also provided explanations for their ratings:
\begin{itemize}
\item [1.] I find that the bot we designed behaves in a way that reflects the specific needs, norms, and culture of our Discord community.
\item [2.] I can easily collaborate with others in bot design using Botender.
\item [3.] I find the test cases helpful in revealing opportunities to improve the bot.
\item [4.] I find the test cases helpful in surfacing situations where people might have differing opinions about whether the bot’s response is appropriate.
\item [5.] I find the experience of designing the bot with the Botender system integrates well with my usage of Discord.
\end{itemize}

Finally, they were asked to suggest features for improvement or for future versions, and to indicate whether they would like to continue using Botender in their server after the study.

\subsection{Study Results}

We present the findings from our field study in the following sections.\footnote{In the field study results, participant IDs are indicated with an S, and the groups they belong to are indicated with a G.} Section \ref{ch6:sec:field_study:tasks} presents the types of tasks participants designed for their community bots and their perceptions of these tasks, as shared in the post-study survey. Our results show that participants were able to design a wide variety of tasks tailored to their communities, and they felt these tasks reflected their unique community needs and culture. Section \ref{ch6:sec:field_study:cases} explores how participants collaborated using case-based provocations. Across the six communities, participants created a total of over 100 proposals and 800 saved test cases during the study. Analysis of the multiple-choice questions that participants answered after each proposal edit revealed they were more likely to iterate on proposals in response to Botender’s case-based provocations, which encouraged collective reflection and discussion about desirable bot behaviors. Feedback from the post-study survey provides further insight into how participants collaboratively improved and discussed bot designs based on opportunities identified through these cases. Finally, Section \ref{ch6:sec:field_study:system} presents additional feedback from participants on aspects of Botender's design, beyond the case-based provocations, that they found particularly helpful for collaborative bot design. These include the overall system workflow, seamless integration with the community platform, and the use of natural language for bot design. Overall, participants found that the bot behaved in ways that aligned with their needs and community norms, attributing this to the collaborative design process enabled by Botender. Over 96\% of participants expressed interest in continuing to use Botender after the study period.

\begin{table}
  \caption{The number of tasks deployed and the proposals and cases created by each group during the field study.}
  \label{table:ch6:field_study_stats}
  \begin{tabular}{cccc}
    \toprule
    Group ID & Tasks & Proposals & Cases \\
    \midrule
    G1 & 18 & 32 & 166 \\
    G2 & 4 & 3 & 14 \\
    G3 & 10 & 23 & 204 \\
    G4 & 17 & 37 & 183 \\
    G5 & 16 & 28 & 165 \\
    G6 & 4 & 14 & 68 \\
    \midrule
    Total & 69 & 137 & 800 \\
    \bottomrule
  \end{tabular}
\end{table}

\begin{table*}
  \caption{Selected tasks from each group reflect the diverse range of tasks they designed for the bot to address the unique needs and norms of their communities. The categories of tasks are assigned according to the taxonomy identified in prior work \cite{seering2018social}, with definitions and examples provided in Section \ref{ch6:sec:related_work:community}. The full list of tasks created by each group is included in Appendix \ref{ch6:sec:field_study_appendix:tasks}.}
  \label{table:ch6:example_tasks}
  \begin{tabular}{clp{16em}p{20em}l}
    \toprule
    Group & Task Name & Trigger & Action & Category\\
    \midrule
    G1 & Sideeyeomatic & Whenever someone says anything questionable or suspicious - things that would generally make someone give them the side eye. & Post this gif: \url{https://tenor.com/p6t9IvV9eBF.gif} & engagement\\
    \midrule
    G2 & Merch Link & Whenever someone asks about or expresses interest in supporting the band, or buying band merchandise or physical copies of the music, or mentions that they enjoy the types of items we sell including vinyl albums, cassette tapes, band shirts, stickers, etc. & Let them know that we have merch items including but not limited to shirts, bandanas, stickers, vinyl albums, cassette tapes and direct them to the website [url] to purchase these and other items & promotion\\
    \midrule
    G3 & Lab location & When someone asks about [lab] location or room number or access info & Reply them with [lab name] ([building code] [room number]), mention that they need to request access through [department acronym] form [service portal url]. Also remind them to get access to the [graduate lounge location] to enjoy free coffee and spend their free time or study. Use proper formatting and emojis & information\\
    \midrule
    G4 & Puppy Training & All users in this server own dogs and like to have fun by roleplaying their dogs talking. Whenever a user imitates their dogs through actions such as barking or voices thoughts from the perspective of their dog, you should trigger & To encourage responsible dog behaviour and also set examples of proper dog behaviour, please praise or scold users as if they are a dog when dogs are mentioned. Users believe their dogs (rightfully so) are very cute, so try to address pets by pet names like "puppy" or "doggy" rather tha scientific terms such as "dog" or "canine" & engagement \\
    \midrule
    G5 & Woah, easy now & Detect angry or aggressive language & Act like a old timey southern cowboy who is trying to calm down his horse. & moderation \\
    \midrule
    G6 & Info overview & Any question about hackathons, [hackathon event], [student club] & Link to [event website] for [hackathon event] specific questions. If asking about what a hackathon is then provide overview of hackathon. If asking about [student club], link to [club url] page as well as provide information about the club. & information \\
    \bottomrule
  \end{tabular}
  
\end{table*}

\subsubsection{\textbf{With Botender, participants designed a variety of tasks for the bot tailored to the specific needs and norms of their own communities.}}\label{ch6:sec:field_study:tasks} Table \ref{table:ch6:field_study_stats} presents descriptive statistics from the field study, including the number of tasks each group deployed and the number of proposals and cases they created to support task deployment. To provide a glimpse of the resulting tasks, Table \ref{table:ch6:example_tasks} presents a sample of tasks deployed by each group. While the current version of Botender only supports the creation of tasks involving single-turn interactions (as mentioned in Section \ref{ch6:sec:botender:system_architecture}), the range of tasks is quite broad and diverse. As one participant noted, these tasks span everything \textit{``from funny quips to actual advice and help to games and welcome messages''} (S4, G5). The tasks participants created for the bot align with prior research \cite{seering2018social}, demonstrating a balance between task-oriented and socially-oriented functions, or combinations of both. As shown in Table \ref{table:ch6:example_tasks}, the types of tasks participants created generally reflect the nature of the server, whether it is more professionally or socially oriented.

Participants agreed with the statement that ``\textit{the bot we designed behaves in a way that reflects the specific needs, norms, and culture of our Discord community}'' ($M = 6.10, SD = 0.94$). They shared that the bot behaves in ways that \textit{``adhere to our culture, humor and jokes''} (S4, G5), and \textit{``reflects us as a community a lot since we approached the tasks we made with a lot of light humor while keeping the usefulness aspect''} (S22, G1). Participants were impressed by how specifically the bot could be tailored to fit their server: \textit{``We came up with some really specific ways to greet our fans. [...] We specifically asked [the bot] to greet people in a way that was both fabulous and gay and fun, but at the same time sad because our band plays sad country songs. It did such a great job of that. It was unbelievable''} (S27, G2). In addition to the content of the messages the bot sends (as defined by a task's action), participants also appreciated the precise timing of when the bot chimes in (as determined by its trigger). For example, a participant found \textit{``the fact that it could be prompted in such detail, made it great with timing [...], which made for a fun moment in the server''} (S15, G4).

\begin{figure}[t!]
  \centering
  \includegraphics[width=\linewidth]{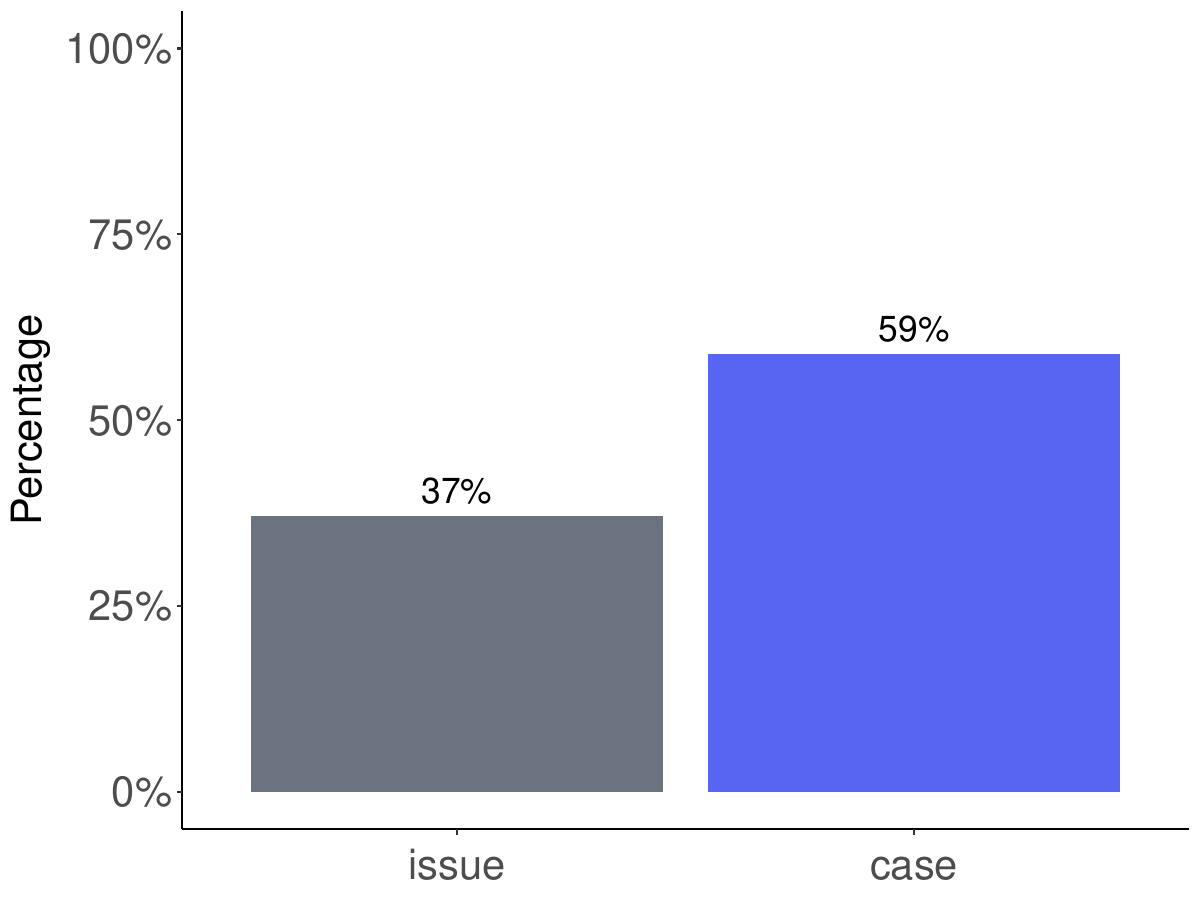} 
  \caption{The percentage of proposal edits aimed at addressing specific cases participants reviewed versus general issues that they thought of during the proposal iteration process. A significantly higher proportion of edits were inspired by specific cases.}
  \Description{A bar chart illustrating the percentage of proposal edits made to address specific cases reviewed by participants versus general issues identified during the proposal iteration process. Edits targeting specific cases accounted for 59\%, while those addressing general issues made up 37\%.}
  \label{fig:ch6:p_percent_edit_questionnaire}
\end{figure}

\subsubsection{\textbf{Participants found that case-based provocations effectively facilitated iterative, collaborative design.}}\label{ch6:sec:field_study:cases} To understand how participants iteratively developed these tasks, we examined the results from the multiple-choice question they answered each time they edited a proposal. Specifically, participants were asked whether their motivation for making an edit was based on a specific case or a general issue. As shown in Figure \ref{fig:ch6:p_percent_edit_questionnaire}, a one-sided hypothesis test revealed that, when using Botender, a larger proportion ($p<0.001$) of proposal edits were aimed at addressing specific cases (59\%) rather than general issues (37\%). Meanwhile, 95\% of the 800 cases saved to the proposals were generated by Botender's case-based provocation algorithm, rather than being manually entered by participants. Taken together, these results suggest that Botender’s case-based provocations effectively facilitate iterative proposal editing during the design process. This aligns with participants’ experiences, as they reported that Botender's case-based provocation \textit{``significantly reduced headaches in prompt engineering and what would have required additional proposals to fix. [...] It allowed me to iterate quickly on my prompts while drafting the task for the proposals. I often didn't even need to make any manual test cases thanks to the generating feature which reduced friction in quickly getting a proposal together''} (S6, G5).

Participants’ self-reported ratings and explanations in the post-study survey further support this finding. Specifically, they agreed that \textit{``they find the test cases helpful in revealing opportunities to improve the bot''} ($M = 6.31, SD = 1.00$) and shared experiences of how the cases supported their iterative process. For example, one participant noted that the test cases \textit{``allowed me to fine tune the description I wrote for letting the bot know when it should respond. This helped me to realize that sometimes I was not writing detailed enough descriptions''} (S22, G1). Another participant mentioned that the \textit{``test cases reveal niche scenarios or unintended uses that may cause the users to reflect and change the original prompt. This was very useful''} (S8, G5). These user experiences align with the goal of Botender’s case-based provocation algorithm, which is designed to uncover common issues often overlooked by non-AI experts, such as ambiguous prompts or unintended downstream consequences, rather than just generating cases that validate what users have already written. As one participant put it: \textit{``The test cases were helpful not just in guiding the bot about false positives/negatives, they also helped clarify the task based on the kinds of test cases it generated''} (S3, G3). Participants appreciated the thought-provoking and diverse test cases, stating \textit{``the test cases were brilliant and diverse. [...] It was helpful especially because it generated some scenarios which I may not have thought of myself to create''} (S14, G6). They also shared that the \textit{``test cases make it very easy to spot when prompting is faulty or needs tweaking. I tweak my prompts at least once for each of my proposals, based off what I see when testing''} (S23, G4) and \textit{``the test cases made it easy to see how a proposal would progress''} (S26, G4).

In addition to supporting design iteration, participants reported that Botender's case-based provocations effectively surfaced differing opinions and sparked discussion about desirable bot behavior within the community. In particular, participants agreed that \textit{``they find the test cases helpful in surfacing situations where people might have differing opinions about whether the bot’s response is appropriate''} ($M = 5.45, SD = 1.40$). Participants found that the test cases \textit{``highlighted gray areas where one person might see the response as appropriate while another might not. For example, in cases involving tone, whether the bot should be more direct or more playful, the test cases made those differences in perspective visible''} (S2, G3). People’s differing reactions to these gray areas \textit{``laid the groundwork for further discussion''} (S25, G4) or \textit{``gave us new alterations or entirely new ideas''} (S13, G5). For example, a participant shared, \textit{``The upvote/downvote system works perfectly for finding which behaviors are controversial. [...] We would notice that there's not a pure consensus on one of the saved test case response[s], which would spark a small discussion''} (S6, G5). However, a few participants mentioned that they didn’t encounter much disagreement \textit{``because we all had similar thoughts''} (S27, G2). Among the groups that participated, we observed that this tended to occur in the smallest communities.

Finally, participants found the test cases helpful in guiding their collective decision-making on whether to deploy a proposal. For example, as one participant shared: \textit{``It's very easy for a chatbot to misunderstand a prompt, the same goes for a human when only looking at what the prompt says. The actual test cases gives a realistic view of what the implementation will look like. We had a task that referenced religion, which was a controversial topic to some, but the test cases served to show that Botender was doing so in an appropriate way and did not step out of line, and that eventually made us agree on implementing the task''} (S15, G4). Similarly, another participant shared that the test cases sometimes led them to decide against deploying a proposal after seeing how the bot would actually respond: \textit{``I saw this with the gaslighting bot, when people in the group saw the responses it generated were in general invalidating of ones feelings, the group was hesitant on approving it''} (S18, G1). Participants found the test cases useful for collective decision-making because \textit{``the test cases give us a good idea on how the bot reacts to different situations, it also shows where it thrives/where its more limited in its capabilities''} (S13, G5).

\subsubsection{\textbf{Participants emphasized that Botender’s overall workflow and integration with their community platform were helpful in fostering community participation.}}\label{ch6:sec:field_study:system} Regarding Botender's overall workflow, participants agreed that \textit{``they can easily collaborate with others in bot design using Botender''} ($M = 6.00, SD = 1.10$). From proposing and iterating on tasks to deploying them, participants found Botender’s workflow effectively facilitated collaboration and ensured that tasks reflected the community’s needs and norms. For example, participants found the proposal design intuitive and well-organized, making it easy to navigate desired changes: \textit{``I think the interface[s] for creating proposals and tasks were very easy to navigate and made it easy to collaborate with others! I liked seeing the entire list view with everyone's proposals [...] It made it easy for me to see which ones I wanted to vote for, and which ones were ready to deploy''} (S22, G1). Participants also found that being able to collaboratively edit proposals helped improve their quality: \textit{``I felt like a lot of my proposals were pretty brief, so I appreciate [another user] editing mine. More than once I noticed he had edited a proposal to add many more details!''} (S19, G1). Meanwhile, participants appreciated the requirement of upvotes for proposal deployment, as it helps ensure that deployed tasks align with the community’s needs and norms: \textit{`[The tasks] fit into the norms and culture of our community. This is evident in the requirement of needing at least 3 upvotes to deploy, meaning that we as a group decided what did and did not mesh`} (S18, G1). In the same vein, the voting requirement also \textit{``prevent[ed] a single person from making a bot that would not serve the community well''} (S7, G5).

Participants also emphasized that deep integration with the community platform they already use is key to keeping community members actively engaged in the bot design process. They agreed that \textit{``the experience of designing the bot with the Botender system integrates well with their usage of Discord''} ($M = 6.14, SD = 1.57$), and shared specific integrations they found particularly valuable. For example, many participants expressed how much they appreciated having a dedicated discussion thread for each proposal on Discord: \textit{``It was an amazing user experience, having it automatically create threads and open/close proposals automatically was very efficient, and felt very modern''} (S11, G6). They believed that \textit{``the built-in notifications kept everyone in the loop. For example, whenever someone made a new proposal, the notification feature made it easy to see updates right away and give feedback without missing anything''} (S2, G3). They found this approach effectively coordinated discussions because \textit{``using threads for the ticketing system eliminates need to manually organize the discussions and keeps everything in one place''} (S6, G5). In addition to the threading design, participants appreciated how details like the login process and overall aesthetics seamlessly fit with their use of Discord. For example, participants shared \textit{``it's very simple to manage the bot via the OAuth login dashboard and we had no issues because of its deep integration with Discord''} (S6, G5), and \textit{``I think this is my favorite part about Botender! [...] it is very easy to use and very much fits the vibes and general aesthetic of Discord very well''} (S15, G4). Overall, participants found \textit{``Botender fit naturally into how I already use Discord [...], which made collaboration smooth and kept the focus on shaping the bot, not juggling tools''} (S2, G3).

Finally, while using natural language to write prompts is not a unique feature of the Botender system, several participants noted that this greatly facilitates broader community participation in bot design. For example, a participant shared that \textit{``the way it’s designed allows me to describe tasks in plain English, and it understands what I mean without needing technical setup. [...] It reduces friction, lets me focus on shaping the community’s culture, and makes it easier to adapt the bot’s behavior to fit our needs. [...] That makes it accessible not only to me but also to other community members who may not have a technical background''} (S2, G3). A participant without coding background echoed this sentiment, sharing that setting up other bots was much more difficult compared to Botender: \textit{``We liked how easy it was to implement ideas. We could talk about them, and then up vote or down vote them based on what we thought. In the past anything like this was really difficult to use. It took a lot of planning and stress. This did not feel that way at all''} (S27, G2). Overall, participants found \textit{``the bot is a great addition to the server. The fact it's so easy to customize makes it really simple for any member to contribute, and shape something that reflects what we needs''} (S25, G4). In the post-study survey, 96\% of participants (30 out of 31) expressed interest in continuing to use Botender after the study.

\section{Discussion}\label{ch6:sec:discussion}

Bots play vital roles and act as essential socio-technical infrastructure within online communities. It is crucial for communities to collaboratively design bots that meet their specific needs and norms, rather than leaving this to outsiders or just a few technically skilled members. In this paper, we present Botender, a system that supports communities in collaboratively proposing, iterating on, and deploying bots powered by LLM-based agents. In particular, Botender facilitates this collaborative design process through case-based provocations, concrete interaction scenarios generated to provoke user reflection and discussion about desirable bot behaviors within their community. We conducted a validation study and a field study to understand how people perceive and use these case-based provocations to collaboratively design bots in real-world communities. Through a validation study (Section \ref{ch6:sec:validation_study}), we saw that participants found Botender's case-based provocations revealed more opportunities for bot improvement, compared with standard test cases. Through a field study (Section \ref{ch6:sec:field_study}), we found that real-world Discord communities effectively collaborated on designing bots tailored to their specific needs and norms using Botender’s case-based provocations. This collaboration was further supported by Botender’s overall collaborative workflow, seamless integration with the community platform, and the use of natural language for bot design.

In this section, we discuss limitations of the current instantiation of Botender that participants wished to overcome, as expressed in their post-study surveys or in their unsuccessful attempts to design bots during the study. For each limitation, we outline opportunities for future HCI research to better support the collaborative and participatory design of community bots powered by AI agents.

\subsection{Advancing Case-Based Provocations}
\subsubsection{\textbf{Limitations}: The current case-based provocation algorithm does not highlight potential conflicts between tasks, and may require more inputs to better surface potential disagreements.}
The case-based provocation algorithm presented in this paper represents a proof-of-concept. Our current case-based provocation algorithm generates three specific types of cases, corresponding to common prompt design pitfalls, such as ambiguous phrasing within prompts, which also represent potential areas of disagreement among collaborators. While most participants found that the current case-based provocations effectively facilitated iterative, collaborative design, they also identified opportunities for improvement. For example, participants expressed a desire to see provocations that reveal situations where two tasks might be in conflict, in order to help \textit{``prevent conflicting task triggers''} (S18, G1). As another example discussed in Section \ref{ch6:sec:field_study:cases}, one participant group (G2) didn't encounter much disagreement \textit{``because we all had similar thoughts''} (S27, G2) and \textit{``everybody seemed to be on the same page''} (S29, G2). While this may simply reflect the dynamics of that particular group, it could also indicate opportunities to improve the case generation algorithm, as we discuss below.

\subsubsection{\textbf{Opportunities:} Future algorithms could generate additional types of case-based provocations and incorporate user feedback on cases to better identify potential areas of disagreement.}
There are several opportunities for future research to explore the design of case generation algorithms that can more effectively provoke user reflection and discussion throughout iterative design processes. For example, building on participants' feedback, future work could explore new types of cases not covered in this study, such as those arising from other common prompt design pitfalls, cases that highlight conflicting tasks, or cases that are at the borderline of acceptability for a \textit{specific community}. We see potential for future case-based provocation algorithms to more effectively surface potential disagreements within a community by learning about divergent perspectives within the community over time, based on users' interactions---such as users' thumbs up/down feedback on individual cases or their discussions about bot behavior (cf.~\cite{koshy2025venire}). More broadly, we see exciting opportunities to extend the concept of case-based provocations beyond prompt design. For example, future research could explore how algorithmically generated cases might be used to provoke reflection, deliberation, and improvements in the design of other kinds of interactive technologies, or the development of public policy~\cite{kuo2025policycraft}.

\subsection{Enhancing Support for Prompt Refinement}
\subsubsection{\textbf{Limitations:} Some participants desired more direct assistance in refining prompts}
With the current version of Botender, users can reflect on and discuss a case generated from a prompt, and then manually refine the prompt as needed. However, some participants anticipated that giving a thumbs up or down to a case would not only surface disagreements among users, but also directly contribute to improving the bot. For example, a participant initially \textit{``expected the bot will learn [from] it and change the response to something else, but it didn't''} (S5, G3). Relatedly, participants expressed a desire to manually correct a case by entering \textit{``the expected output to make the bot learn''} (S5, G3), or to \textit{``directly explain to the bot why a case is good and why a case is bad, so it would be able to improve''} (S7, G5). This feedback highlights a desire for more assistance in prompt refinement, potentially based on users' feedback on cases, rather than leaving the refinement process completely manual.

\subsubsection{\textbf{Opportunities:} Future systems could offer prompt refinement suggestions based on a variety of human feedback on cases.}
Building upon participants' feedback, future systems could consider offering more ways for users to provide feedback on cases, such as allowing them to correct the bot's response or provide direct explanations \cite{mcdaniel1999getting}, in addition to giving a thumbs up or down. The system could then incorporate this human feedback to suggest refined prompts, which users would have the agency to adopt, revise, or ignore as they see fit. While a fully automated prompt refinement process, where the bot evolves on its own without users having access to its prompt, may seem simpler and perhaps more desirable, prior HCI research suggests otherwise \cite{smith2020keeping}. Communities often prefer to retain agency and control over the AI systems they use or rely on, rather than having this control completely hidden within the system or managed solely by outsiders \cite{kuo2024wikibench, halfaker2020ores}. Moreover, communities' needs and preferences may change over time, meaning that objectives set in the past may no longer align with their current goals \cite{halfaker2025collective}. For these reasons, we recommend that future work seek a balance, to keep the community actively involved and at the center of the design process.

\subsection{Expanding the Capabilities of AI Agents}
\subsubsection{\textbf{Limitations}: Participants wanted to customize tasks that required bots to handle more than single-turn conversations.}
As the first attempt to support collaborative design of AI agents in community contexts, the current version of Botender focuses on enabling the design of single-turn, LLM-based conversational AI agents. This means that the bot and its underlying agents can process only one user message at a time and respond with a single message. Channel names are the only context information that allows the agents to tailor their behavior more specifically. Although participants were aware of this limitation from the beginning of the study, many expressed in the post-study survey a desire for expanded bot capabilities that would allow for greater customization to better address the unique needs of their communities. For example, participants shared feedback such as, \textit{``I would love for Botender to be able to take into context the full conversations!''} (S15, G4), and \textit{``Being able to store data would be neat! For example, setting up a birthday reminder would require saving data about the different users''} (S19, G1). They also expressed interest in additional bot actions beyond single-turn replies, such as \textit{``the ability to block messages''} (S9, G6).

\subsubsection{\textbf{Opportunities}: Future bots could build upon Botender's system architecture to expand their capabilities for greater customization.}
As mentioned in Section \ref{ch6:sec:botender:system_architecture}, Botender's system and agent architecture is designed to handle a broader range of platform events, actions, and context information. On the event side, in addition to listening to user messages, Botender's listener can detect a wider range of platform events provided by the platform API, such as users joining channels, the creation of new threads, or changes to user permissions. These events can be translated into information that agents can interpret and use to provide appropriate action instructions. Similarly, on the action side, Botender's action executor can carry out a broader range of actions available through the API, such as creating new channels, muting users when appropriate, or searching the internet for up-to-date information.\footnote{We experimented with internet search, but found that it made case generation too slow and affected the user experience. It may become faster as LLMs improve.} The context information available to agents can also be expanded depending on the event. For example, for the event of a user sending a message, the context could include the time, the user’s permission, a greater history of previous messages, or stored data about the community \cite{wang2025social}. Future systems could provide agents with richer context and expand their capabilities based on what would be most useful to communities. However, as prior research points out \cite{hwang2024adopting}, granting bots more permissions can inevitably raise concerns among community members about potential misuse and the risk of unrecoverable consequences. Striking the right balance between agent capabilities and user concerns will be an important challenge to address.

\subsection{Scaling to Broaden Participation}
\subsubsection{\textbf{Limitations:} How to effectively support scaling as the number of users increases remains an open question.}
In our current field study, groups have a maximum of six participants, with deployments on servers of up to 429 members (Table \ref{table:ch6:participants}). Each group has created a maximum of 37 proposals, and the highest number of tasks deployed per group is 18 (Table \ref{table:ch6:field_study_stats}). While this represents a reasonable size for many small to medium Discord communities, there are also much larger communities, such as the Discord server of the Wikimedia Community, which has nearly 10,000 members. Given Wikimedia’s emphasis on broad participation, many of these members may also be interested in contributing to collaborative bot design. As one participant noted, while the current design \textit{``just fits right in, but if you scale up the amount of users making proposals I imagine it could be quite tough''} (S4, G5).

\subsubsection{\textbf{Opportunities:} Future research could explore what design changes or alternatives might better support broader community participation.}
There are several directions that could be pursued. For example, providing clearer guidance on the division of labor could be helpful. Members familiar with community norms could focus on iterating on proposals, while others could contribute by creating cases to support design iterations. There are also opportunities to lower the barrier to collecting feedback on cases. As one participant suggested, \textit{``allowing users to react to a [bot's] message [directly on Discord, rather than on Botender's web interface], could ease the collection of test cases for editing proposals to better the bot's responses''} (S6, G5). Collective decision-making around proposal deployment will also need to be adapted. For instance, it is important to strike a balance between keeping the requirements for deployment accessible while ensuring that implemented changes truly reflect the broader community’s needs. Overall, these challenges relate to broader HCI research on designing systems for community participation, specifically navigating the tradeoff between lowering participation barriers and supporting effective collective action \cite{shaw2014computer, salehi2015we, kuo2024wikibench}.

\subsection{Applying to Different Communities}
\subsubsection{\textbf{Limitations:} Botender's collaborative approach to bot design may not be suitable for all communities with different norms.}
Botender is designed to support a collaborative approach to bot design, involving participation and deliberation among community members. However, it is important to note that the system alone can hardly overcome the power dynamics that may inevitably exist within different communities. For example, in smaller, close-knit groups, such as those in our study, power dynamics may be more distributed, allowing members to freely propose, iterate on, and deploy desired changes to their bots. However, in communities with more hierarchical power structures, where community leaders prefer to design the bot themselves, the applicability of Botender becomes more complex. On one hand, such communities often require heavier moderation and could benefit from Botender's tailored moderation capabilities. On the other hand, the collaborative features of Botender may be less effective, as members may have fewer opportunities to participate in the design process. This complexity was illustrated by a participant who noted: \textit{``The bot would probably be best suited for larger, more rowdy servers, which already struggle with finding competent moderators.''} At the same time, the same participant expressed the desire that \textit{``As someone with higher administrative rights than other members in the server, I think that I should be able to remove a task without needing other people to upvote a proposal''} (S16, G4). Notably, this participant was also the only one (out of 31) who did not express interest in continuing to use Botender after the study.

\subsubsection{\textbf{Opportunities:} Future research could explore ways to support more collaborative and democratic approaches to bot design across different communities.}
Botender is \textit{purposefully} designed to broaden community participation and collaboration in bot design. In fact, the dilemma faced by the participant mentioned above (S16) can be resolved, for example, by setting the deployment threshold of a proposal to one and restricting edit permissions to a single admin. With this configuration, Botender can accommodate this participant's preference by enabling tailored moderation while also centralizing decision-making power in a single individual. However, we have intentionally chosen to pursue a more democratic vision of bot design, and to overcome the unique challenges involved in making this vision a reality. The main design of Botender---including core components like case-based provocations to encourage collective reflection and discussion, along with collaborative features such as discussion threads for each proposal---has all been deliberatively designed to support this vision. We hope that Botender provides community members with an option for a more collaborative, bottom-up approach to bot design, and that our work inspires further research and systems that enable more democratic approaches to community governance \cite{zhang2020policykit, kuo2025policycraft, ovadyaposition, feng2025sociotechnical, kuo2024wikibench}.

\section{Conclusion}
In this work, we have demonstrated how a system can support users in collaborative bot design through case-based provocations. Our findings show that these provocations can effectively surface opportunities for bot improvement, reveal potential sources of disagreement, and support the collaborative bot design process in real online communities. Building on this work, future HCI systems should explore expanding bot capabilities to meet diverse community needs, explore the design of more advanced case-based provocation techniques, address scaling challenges to enable broader participation, and navigate power dynamics within communities.

\begin{acks}

The funding for this research was provided by CMU's Block Center for Technology and Society, Metagov's Grant for Interoperable Deliberative Tools, the National Institute of Standards and Technology (NIST) (ror.org/05xpvk416) and the CMU (ror.org/05x2bcf33) AI Measurement Science and Engineering Center. Tzu-Sheng Kuo (ORCID: 0000-0002-1504-7640) was funded by the K\&L Gates Presidential Fellowship in Ethics and Computational Technologies and NIST through Federal Award ID Number 60NANB24D231. We thank Michael Bernstein, Aniket Kittur, Sherry Tongshuang Wu, Nikolas Martelaro, Chien-Sheng Jason Wu, Pranav Khadpe, K. J. Kevin Feng, Xingyu Bruce Liu, and Tiffany Chih for their insightful feedback on the system design. We are also grateful to Aileen Benedict, Sameer Patil, Estelle Smith, Rotem Guttman, and Joon Jang for their assistance with recruitment for the study. Finally, we thank Isadora Krsek for designing the Botender logo.
\end{acks}

\bibliographystyle{ACM-Reference-Format}
\bibliography{reference}

\appendix
\section{Prompts Used in the Botender System}\label{ch6:sec:botender_appendix}
In this section, we provide all the prompts used for the LLM in the Botender system, including prompts for the AI agents that power the bot and the LLM module used in Botender's case-based provocation algorithm. \textit{System prompts} refer to the instructions given to each LLM, while \textit{user prompts} are the inputs provided to the LLM to generate responses.

\subsection{Prompts for the AI Agents}\label{ch6:sec:botender_appendix:agents}
\subsubsection{Orchestrator Agent}

\subparagraph{\footnotesize \texttt{\textbf{System Prompt}:\\
You are a helpful assistant tasked with determining whether a task should be triggered based on a user's message in a specific channel. You will receive a list of tasks, each with an associated ID and trigger condition, as well as the user's message and the channel where it was sent. If the message is relevant to the trigger condition of a specific task, respond with that task's ID. If the message is relevant to multiple tasks, respond with the ID of the task to which it is most relevant. If the message does not match any task trigger, respond with 0. Your response must be a JSON object with a single key "taskId". For example: \{"taskId": "some-task-id"\} or \{"taskId": "0"\}.\\
}}

\subparagraph{\footnotesize \texttt{\textbf{User Prompt}:\\
Here is a list of tasks:\\\\
Task ID: [taskId]\\
Task Trigger: [task trigger]\\
\vdots\\
Task ID: [taskId]\\
Task Trigger: [task trigger]\\\\
User message in the \#[channel] channel: [user message]\\
}}

\subsubsection{Task-Specific Agent}

\subparagraph{\footnotesize \texttt{\textbf{System Prompt}:\\
You are a helpful assistant tasked with responding to a user's message in a specific channel, following the instructions provided in an assigned action. You will receive the action instructions, the user's message, and the channel where it was sent. Based on the action, compose an appropriate reply. If you determine that no response is necessary, use "n/a". Your response must be a JSON object with a single key "response". For example: \{"response": "Here is your reply."\} or \{"response": "n/a"\}.\\
}}

\subparagraph{\footnotesize \texttt{\textbf{User Prompt}:\\
Action: [task action].\\
User message in the \#[channel] channel: [user message]\\
}}

\subsection{Prompts for the Case-Based Provocations}\label{ch6:sec:botender_appendix:algorithm}
As shown in Figure \ref{fig:ch6:pipelines}, Botender's case-based provocation algorithm consists of three LLM pipelines. Each pipeline includes a \textit{detector}, \textit{generator}, and \textit{evaluator} for creating a specific type of provocative case. The prompts for all nine of these LLMs, as well as the final \textit{selector}, are provided in this section. Note that some shared system prompts are repeated across different LLMs, including the instruction limiting the bot's capability to single-turn conversation, the input specification indicating that the bot receives a channel and user message as input, and the community description. The community description describes the general tone of the community and can be customized to make cases more relevant to a specific group, although none of the participant groups chose to modify the default description. All of the shared system prompts are provided at the end.

\subsubsection{Pipeline for Revealing Ambiguous Phrases}

\subparagraph{\footnotesize \texttt{\textbf{Detector System Prompt}:\\
You are a helpful assistant tasked with identifying critical ambiguities in prompts written for language model-based bots deployed within an online community. This prompt defines:\\
\textbullet\quad A trigger: when the bot should take action.\\
\textbullet\quad An action: what the bot should do when triggered.\\
< bot capability >\\
Your Task:\\
Read the full prompt carefully. Identify specific phrases or instructions that are ambiguous, underspecified, and open to multiple reasonable interpretations. Focus exclusively on ambiguities that could cause:\\
\textbullet\quad Vague or undefined concepts\\
\textbullet\quad Unclear boundaries or thresholds\\
\textbullet\quad Conflicting or competing goals\\
\textbullet\quad Situational or contextual assumptions\\
\textbullet\quad Ambiguity about what, when, or how the bot is supposed to act\\
Prioritize ambiguities that could lead to reasonable differences in human interpretation, especially those where people might disagree about whether the bot's behavior is desirable. Focus on ambiguities that could cause visible inconsistencies in the bot's behavior. Do not list trivial ambiguities, style differences, or issues that would not affect how real users experience the bot.\\
Output Format:\\
Return a JSON object containing an array of ambiguities. Each ambiguity should have a unique key starting from 0 and include the following two properties:\\
\textbullet\quad underspecified\_phrase: a specific quote or snippet from the prompt that is ambiguous\\
\textbullet\quad description: a 1-2 sentence explanation of what makes it ambiguous or open to multiple interpretations\\
All values must be JSON-safe: wrap any field that contains commas in quotes, and avoid newlines. Do not include any extra text, formatting, or commentary outside the JSON object.\\
}}

\subparagraph{\footnotesize \texttt{\textbf{Detector User Prompt}:\\
Prompt:\\
\textbullet\quad Trigger: [task trigger]\\
\textbullet\quad Action: [task action]\\
}}

\subparagraph{\footnotesize \texttt{\textbf{Generator System Prompt}:\\
You are a helpful assistant tasked with generating input test cases that explore how ambiguous phrases in a bot's prompt could be interpreted in different, plausible ways. This prompt defines:\\
\textbullet\quad A trigger: when the bot should take action.\\
\textbullet\quad An action: what the bot should do when triggered.\\
< bot capability >\\
You will be provided with:\\
\textbullet\quad prompt: the full prompt for the bot, containing one or more ambiguous phrases.\\
\textbullet\quad underspecified\_phrase: a specific snippet from the prompt that is ambiguous.\\
\textbullet\quad description: a 1-2 sentence explanation describing why the phrase is ambiguous or can be interpreted in multiple ways.\\
Your Task:\\
For each underspecified\_phrase, generate a small set of test cases that illustrate distinct, plausible alternative interpretations of the phrase. A test case is an input to the bot that adheres to the following input specification:\\
< input specification >\\
When generating test cases, prioritize those that provoke visible divergence in bot behavior—either in whether the bot responds (trigger ambiguity) or in how the bot responds (action ambiguity). Aim to create test cases that illustrate non-obvious yet reasonable interpretations, revealing hidden assumptions, unclear boundaries, or conflicting objectives within the original, underspecified phrase. If the ambiguity influences the bot's action, design the test case to elicit a bot response that clearly diverges from its typical default response. If the ambiguity concerns the trigger, focus on whether the bot responds or not. Each test case should make the ambiguity evident at the surface level, discernible from the channel, user message, and bot response alone, without the need for additional explanation.\\
Additionally, the test cases should be realistic and natural, mirroring the typical messages found in the following community and reflecting its unique tone:\\
< community description >\\
Do not generate test cases based on literal, overly obvious, or superficial interpretations. Avoid creating test cases that only involve minor tone or style differences, unless these differences have a clear impact on user-facing behavior. Additionally, do not include cases that would not affect how humans perceive or interact with the bot.\\
Output Format:\\
Return a JSON object containing an array of the generated test cases. Each case should have a unique key starting from 0 and include the following four properties.\\
\textbullet\quad underspecified\_phrase: the specific snippet from the prompt that is ambiguous.\\
\textbullet\quad interpretation: a plausible alternative interpretation of the phrase that the test case is generated to illustrate.\\
\textbullet\quad reasoning: a brief explanation of how the test case reveals the ambiguity.\\
\textbullet\quad case: the input test case, formatted according to the input specification.\\
All values must be JSON-safe: wrap any field that contains commas in quotes, and avoid newlines. Do not include any extra text, formatting, or commentary outside the JSON object.\\
}}

\subparagraph{\footnotesize \texttt{\textbf{Generator User Prompt}:\\
prompt:\\
\textbullet\quad Trigger: [task trigger]\\
\textbullet\quad Action: [task action]\\
underspecified\_phrase: [underspecified\_phrase from the detector’s output]\\
description: [description from the detector’s output]\\
}}

\subparagraph{\footnotesize \texttt{\textbf{Evaluator System Prompt}:\\
You are a helpful assistant tasked with evaluating whether a test case clearly demonstrates a plausible and critical alternative interpretation of an ambiguous phrase in a bot's prompt. This prompt defines:\\
\textbullet\quad A trigger: when the bot should take action.\\
\textbullet\quad An action: what the bot should do when triggered.\\
< bot capability >\\
You will be provided with:\\
\textbullet\quad prompt: the full prompt for the bot, including both the trigger and action components.\\
\textbullet\quad underspecified\_phrase: a specific snippet from the prompt that is ambiguous.\\
\textbullet\quad interpretation: a plausible alternative interpretation of the phrase that the test case is intended to illustrate.\\
\textbullet\quad reasoning: a brief explanation describing how the test case could demonstrate this interpretation.\\
\textbullet\quad case: the test case itself, including the user message in a specific channel, the specific task triggered for the bot (if any), and the corresponding bot response to that task.\\
It is possible that the user input does not trigger any task, or that the bot chooses not to respond even if a task is triggered.\\
Your Task:\\
Decide whether the test case clearly and directly demonstrates the intended interpretation based only on the channel, user message, and bot response. The ambiguity must be apparent to a human without explanation. Only approve the case if it clearly tests the goal stated in the input reasoning.\\
At the same time, reject any test cases where the scenario assumes the bot can perform actions beyond its defined capabilities. Also, reject cases where the interpretation shown is non-critical—that is, it does not impact user understanding or the bot's behavior. Additionally, reject test cases that simply reflect an expected, default, or literal reading of the ambiguous phrase, as well as those where the demonstrated interpretation is too subtle for an average human to notice.\\
If the ambiguity involves how the bot should respond—meaning the action within the prompt is underspecified—consider the following additional steps: First, infer the generalized or default response the bot would typically give based on the prompt and input. Next, compare this default response to the bot's actual response in the test case. Approve the case only if the actual response shows a clear and noticeable difference from the default in terms of tone, structure, or content, such that the change would be obvious to a human observer. Minor shifts in tone, phrasing, or politeness do not count unless they lead to a significant change in the bot's observable behavior.\\
Output Format:\\
Return a JSON object with the following two properties:\\
\textbullet\quad label: a boolean value—true if the test case visibly and meaningfully demonstrates the intended interpretation of the underspecified phrase; false if it does not, or if it is rejected.\\
\textbullet\quad label\_explanation: a brief, 1 to 2 sentence explanation supporting your decision.\\
All values must be JSON-safe: wrap any field that contains commas in quotes, and avoid newlines. Do not include any extra text, formatting, or commentary outside the JSON object.\\
}}

\subparagraph{\footnotesize \texttt{\textbf{Evaluator User Prompt}:\\
prompt:\\
\textbullet\quad Trigger: [task trigger]\\
\textbullet\quad Action: [task action]\\
underspecified\_phrase: [underspecified\_phrase from the detector’s output]\\
interpretation: [interpretation from the generator's output]\\
reasoning: [reasoning from the generator's output]\\
case:\\
\textbullet\quad channel: [channel from the generator's output]\\
\textbullet\quad user message: [user message from the generator's output]\\
\textbullet\quad triggered task: [triggered task from the bot]\\
\textbullet\quad bot response: [bot response from the bot]\\
}}

\subsubsection{Pipeline for Revealing Overly Narrow Phrases}
\subparagraph{\footnotesize \texttt{\textbf{Detector System Prompt}:\\
You are a helpful assistant tasked with identifying critical overspecified phrases in prompts written for language model-based bots. This prompt defines:\\
\textbullet\quad A trigger: when the bot should take action.\\
\textbullet\quad An action: what the bot should do when triggered.\\
< bot capability >\\
Your Task:\\
Read the full prompt carefully. Identify overspecified phrases—parts of the prompt that unnecessarily limit the bot's behavior or responses, phrased too narrowly, rigidly, or tied to surface-level specifics. These may prevent the bot from fulfilling its broader functional purpose.\\
Follow these steps to complete your task:\\
1. Infer the Broader Goal: Read the full prompt carefully. Infer the broadest reasonable functional goal: what the bot is ultimately intended to detect, prevent, encourage, or support, independent of any surface-level constraints or examples mentioned in the wording of the prompt. Focus on the underlying user problem, situation, or need that the bot is designed to address. Ignore specific conditions, instances, or implementation details unless they are essential to the bot’s purpose. Express the broader goal as what the bot should ideally support, if it were not constrained by unnecessary restrictions.\\
2. Identify Overspecified Phrases: Identify specific snippets of the prompt that unnecessarily constrain how the bot can fulfill its broader goal. Focus on requirements tied to particular content types, formats, channels, or contexts; examples treated as strict conditions; and narrow definitions that exclude plausible situations fitting the broader goal.\\
3. Define Uncovered Scenarios: For each overspecified phrase, describe as thoroughly as possible the set of scenarios that are currently excluded because of the restrictive wording. These scenarios should fit within the broader goal and could reasonably be handled by the bot without requiring any expansion of its capabilities.\\
Important: Do not include scenarios that are already covered by the current overspecified phrase. Think of uncovered scenarios as the portion of the broader goal left unaddressed due to the overspecified phrase. Apply deliberate creativity: consider realistic, plausible situations that are missed due to unnecessary specificity. Focus on diverse, meaningful cases that reflect the variety of user needs the bot is intended to support. Prioritize scenarios that are plausible within the community where the bot is deployed, likely to arise in typical use, and distinct from one another in form, context, or content.\\
Output Format:\\
Return a JSON object containing an array of overspecified phrases. Each phrase should have a unique key starting from 0 and include:\\
\textbullet\quad broader\_goal: the broader goal of the prompt, as you inferred from its content.\\
\textbullet\quad overspecified\_phrase: a specific quote or snippet from the prompt that is overly specific.\\
\textbullet\quad uncovered\_scenarios: a description of scenarios that are relevant to the broader goal but are not addressed by the current overspecified phrase.\\
All values must be JSON-safe: wrap any field that contains commas in quotes, and avoid newlines. Do not include any extra text, formatting, or commentary outside the JSON object.\\
}}

\subparagraph{\footnotesize \texttt{\textbf{Detector User Prompt}:\\
Prompt:\\
\textbullet\quad Trigger: [task trigger]\\
\textbullet\quad Action: [task action]\\
}}

\subparagraph{\footnotesize \texttt{\textbf{Generator System Prompt}:\\
You are a helpful assistant tasked with generating input test cases that illustrate how an overspecified phrase in a prompt might cause the bot to miss relevant situations. This prompt defines:\\
\textbullet\quad A trigger: when the bot should take action.\\
\textbullet\quad An action: what the bot should do when triggered.\\
< bot capability >\\
You will be provided with:\\
\textbullet\quad prompt: the full prompt for the bot, containing one or more overspecified phrases.\\
\textbullet\quad overspecified\_phrase: a specific snippet from the prompt identified as overly specific.\\
\textbullet\quad broader\_goal: the broader goal of the prompt.\\
\textbullet\quad uncovered\_scenarios: a description of scenarios that are relevant to the broader goal but excluded by the overspecified phrase.\\
Your Task:\\
For each overspecified\_phrase, generate distinct test cases, where each case directly reflects one specific uncovered scenario from the provided list, aligns with the broader goal, and is currently excluded due to the overspecified phrase. A test case is an input to the bot that adheres to the following input specification:\\
< input specification >\\
Each test case should visibly demonstrate how the overspecified phrase restricts the bot's behavior, excluding relevant situations that fit the broader goal. The missed scenario should be evident from the channel name and user message alone, without requiring further explanation. When designing test cases, prioritize those that surface differences in message content, phrasing, or context that realistically reflect how the overspecified phrase causes the bot to fail. Avoid trivial variations or unrealistic phrasing.\\
Additionally, the test cases should be realistic and natural, mirroring the typical messages found in the following community and reflecting its unique tone:\\
< community description >\\
Do not generate scenarios already covered by the overspecified phrase. Do not generate cases that require capabilities the bot does not have. Do not include trivial, repetitive, or unrealistic cases. The uncovered scenario should be clear to a human reviewer from the input alone.\\
Output Format:\\
Return a JSON object containing an array of generated test cases. Each case should have a unique key starting from 0 and include:\\
\textbullet\quad uncovered\_scenario: the specific uncovered scenario that the test case is generated to illustrate.\\
\textbullet\quad reasoning: a brief explanation describing how the test case makes this uncovered scenario visible to a human reviewer.\\
\textbullet\quad case: the input test case, formatted according to the input specification.\\
All values must be JSON-safe: wrap any field that contains commas in quotes, and avoid newlines. Do not include any extra text, formatting, or commentary outside the JSON object.\\
}}

\subparagraph{\footnotesize \texttt{\textbf{Generator User Prompt}:\\
prompt: \\
\textbullet\quad Trigger: [task trigger]\\
\textbullet\quad Action: [task action]\\
overspecified\_phrase: [overspecified\_phrase from the detector’s output]\\
broader\_goal: [broader\_goal from the detector’s output]\\
uncovered\_scenarios: [uncovered\_scenarios from the detector’s output]\\
}}

\subparagraph{\footnotesize \texttt{\textbf{Evaluator System Prompt}:\\
You are a helpful assistant tasked with evaluating whether a test case effectively demonstrates an uncovered scenario caused by an overspecified phrase in a bot's prompt. This prompt defines:\\
\textbullet\quad A trigger: when the bot should take action.\\
\textbullet\quad An action: what the bot should do when triggered.\\
< bot capability >\\
You will be provided with:\\
\textbullet\quad prompt: the full prompt for the bot, including both the trigger and action components.\\
\textbullet\quad overspecified\_phrase: a snippet from the prompt that is identified as overly specific.\\
\textbullet\quad broader\_goal: the broader goal of the prompt.\\
\textbullet\quad uncovered\_scenario: the scenario the test case is designed to illustrate.\\
\textbullet\quad reasoning: an explanation of how the test case illustrates the scenario that is uncovered by the overly specific phrase in the prompt.\\
\textbullet\quad case: the test case itself, including the user message in a specific channel, the specific task triggered for the bot (if any), and the corresponding bot response to that task.\\
It is possible that the user input does not trigger any task, or that the bot chooses not to respond even if a task is triggered.\\
Your Task:\\
Decide whether the test case clearly and directly demonstrates the uncovered scenario caused by the overspecified phrase. Approve the test case only if it visibly reveals the restriction introduced by the overspecified phrase, showing that the bot fails to address a situation that clearly fits within the broader goal. The scenario must be plausible, relevant to the broader goal, and clearly observable based solely on the input message and bot response.  Approve only when a human reviewer could reasonably understand, from the input message and bot response alone, how the overspecified phrase prevents the bot from acting as intended. Only approve the case if it clearly tests the goal stated in the input reasoning.\\
Reject any test case where the uncovered scenario is unclear, irrelevant, trivial, or not apparent from the case itself. Additionally, reject any test case where the scenario requires the bot to perform actions beyond its defined capabilities.\\
Output Format:\\
Return a JSON object with the following two properties:\\
\textbullet\quad label: A boolean value—true if the test case clearly demonstrates the uncovered scenario; false if it does not, or if it is rejected.\\
\textbullet\quad label\_explanation: a brief, 1 to 2 sentence explanation supporting your decision.\\
All values must be JSON-safe: wrap any field that contains commas in quotes, and avoid newlines. Do not include any extra text, formatting, or commentary outside the JSON object.\\
}}

\subparagraph{\footnotesize \texttt{\textbf{Evaluator User Prompt}:\\
prompt:\\
\textbullet\quad Trigger: [task trigger]\\
\textbullet\quad Action: [task action]\\
overspecified\_phrase: [overspecified\_phrase from the detector’s output]\\
broader\_goal: [broader\_goal from the detector’s output]\\
uncovered\_scenario: [uncovered\_scenarios from the generator’s output]\\
reasoning: [reasoning from the generator’s output]\\
case:\\
\textbullet\quad channel: [channel from the generator's output]\\
\textbullet\quad user message: [user message from the generator's output]\\
\textbullet\quad trigger task: [triggered task from the bot]\\
\textbullet\quad bot response: [bot response from the bot]\\
}}

\subsubsection{Pipeline for Revealing Consequential Phrases}
\subparagraph{\footnotesize \texttt{\textbf{Detector System Prompt}:\\
You are a helpful assistant tasked with identifying potential unintended consequences in prompts written for language model-based bots deployed within an online community. This prompt defines:\\
\textbullet\quad A trigger: when the bot should take action.\\
\textbullet\quad An action: what the bot should do when triggered.\\
< bot capability >\\
Your Task:\\
Read the full prompt carefully. Identify specific phrases or instructions that could lead to unintended community-level consequences. Focus on aspects of the prompt that may produce negative impacts on participation, trust, tone, or community experience—even if the prompt appears clear or well-intentioned. Surface potential value tensions, social risks, and moderation pitfalls that the community may wish to proactively consider or address. Focus on raising concerns about the prompt's direction, tone, or broader social implications, rather than evaluating its precision or scope. Your goal is to help the community clarify its values and anticipate potential risks before deployment.\\
Draw from the following four types of potential unintended consequences of the bot to guide your analysis. These consequences are especially useful for prompting community reflection, surfacing implicit values, and encouraging more thoughtful moderation design:\\
1. Encouraging Contribution: Bots may unintentionally discourage participation by overemphasizing metrics or feedback, crowding out users' intrinsic motivation to learn, explore, or contribute creatively. Praise or corrections may feel impersonal or manipulative if delivered rigidly by a bot, undermining trust and commitment. Bots may also reinforce dominant behaviors or popular contributions, marginalizing diverse or alternative forms of value. Replacing personal recognition with automated responses may erode the human connection essential for healthy participation.\\
2. Encouraging Commitment: Bots that overlook users' prior efforts, personal goals, or community identity signals may weaken ongoing participation. Ignoring users' history of contributions, social ties, or personal motivations (like fun or growth) can reduce their investment in the community. Overly procedural enforcement may disrupt the sense of belonging and shared identity that helps retain contributors.\\
3. Regulating Behavior: Bots may enforce norms in ways that feel confusing, unfair, or alienating. Responses may lack clarity or consistency, punish users without giving them a dignified way to recover, or impose overly harsh or arbitrary sanctions that erode trust. Automated moderation risks appearing punitive rather than supportive, especially if responses feel generic or opaque. Failing to track repeat issues or ignoring community tone can further damage perceptions of fairness, legitimacy, and ownership.\\
4. Managing Newcomer Integration: Newcomers may be deterred if bots apply strict rules too early, fail to explain expectations clearly, or do not provide enough early guidance. Rigid enforcement or unclear onboarding may lead to confusion, early mistakes, and disengagement. Bots that present norms too formally or too casually may mislead newcomers about the community's actual tone or values. Abrupt exposure to complex tasks without scaffolding may overwhelm or alienate new participants.\\
Prioritize unintended consequences of the prompt that could significantly affect real user experience. The unintended consequence you identify should be something that can be addressed by revising the prompt's wording, without needing to expand the bot's capabilities. Avoid trivial issues, style preferences, or theoretical edge cases unlikely to occur in practice.\\
Output Format:\\
Return a JSON object containing an array of potential unintended consequences. Each consequence should have a unique key starting from 0 and include the following two properties:\\
\textbullet\quad problematic\_phrase: a specific quote or snippet from the prompt that could potentially cause unintended consequences.\\
\textbullet\quad consequence: a 1 to 2 sentence explanation of the possible unintended consequence or concern related to this phrase\\
All values must be JSON-safe: wrap any field that contains commas in quotes, and avoid newlines. Do not include any extra text, formatting, or commentary outside the JSON object.\\
}}

\subparagraph{\footnotesize \texttt{\textbf{Detector User Prompt}:\\
Prompt:\\
\textbullet\quad Trigger: [task trigger]\\
\textbullet\quad Action: [task action]\\
}}

\subparagraph{\footnotesize \texttt{\textbf{Generator System Prompt}:\\
You are a helpful assistant tasked with generating input test cases that illustrate how specific problematic phrases in a language model-based bot's prompt could unintentionally cause harm to the online community where the bot is deployed. These test cases are intended to reveal how the bot's current design may challenge important community values and spark thoughtful reflection on the behaviors the community wishes to encourage.\\
The prompt of the bot defines:\\
\textbullet\quad A trigger: when the bot should take action.\\
\textbullet\quad An action: what the bot should do when triggered.\\
< bot capability >\\
You will be provided with:\\
\textbullet\quad prompt: the full prompt for the bot, containing one or more potentially problematic phrases.\\
\textbullet\quad problematic\_phrase: a specific snippet from the prompt that could potentially cause unintended consequences.\\
\textbullet\quad consequence: the possible unintended consequence identified as a result of the potentially problematic phrase.\\
Your Task:\\
For each identified consequence, create a single, credible test case that naturally depicts how this consequence might arise. A test case is an input to the bot that adheres to the following input specification:\\
< input specification >\\
Each test case should stand alone as a compelling, credible example—illustrating the tension between the prompt and the community value at risk. The consequence should be visible at the surface level, without relying on further explanation.\\
Additionally, the test cases should be realistic and natural, mirroring the typical messages found in the following community and reflecting its unique tone:\\
< community description >\\
Output Format:\\
Return a JSON object with the following two properties:\\
\textbullet\quad reasoning: a brief explanation of how the test case reveals the unintended consequence.\\
\textbullet\quad case: the input test case, formatted according to the input specification.\\
All values must be JSON-safe: wrap any field that contains commas in quotes, and avoid newlines. Do not include any extra text, formatting, or commentary outside the JSON object.\\
}}

\subparagraph{\footnotesize \texttt{\textbf{Generator User Prompt}:\\
prompt:\\
\textbullet\quad Trigger: [task trigger]\\
\textbullet\quad Action: [task action]\\
problematic\_phrase: [problematic\_phrase from the detector’s output]\\
consequence: [consequence from the detector’s output]\\
}}

\subparagraph{\footnotesize \texttt{\textbf{Evaluator System Prompt}:\\
You are a helpful assistant tasked with evaluating whether a test case clearly demonstrates how a specific problematic phrase in a language model-based bot's prompt could lead to unintended negative consequences for the online community where the bot is deployed. The bot's prompt defines:\\
\textbullet\quad A trigger: when the bot should take action.\\
\textbullet\quad An action: what the bot should do when triggered.\\
< bot capability >\\
You will be provided with:\\
\textbullet\quad prompt: the full prompt for the bot, including both the trigger and action components.\\
\textbullet\quad problematic\_phrase: a specific snippet from the prompt that could potentially cause unintended consequences.\\
\textbullet\quad consequence: the possible unintended consequence identified as a result of the potentially problematic phrase.\\
\textbullet\quad reasoning: a brief explanation of how the test case reveals the unintended consequence.\\
\textbullet\quad case: the test case itself, including the user message in a specific channel, the specific task triggered for the bot (if any), and the corresponding bot response to that task.\\
It is possible that the user input does not trigger any task, or that the bot chooses not to respond even if a task is triggered.\\
Your Task:\\
Decide whether the test case clearly and convincingly demonstrates the described unintended consequence. Approve the test case only if the consequence is visibly illustrated through the input and bot response (if any), the scenario is realistic, relevant to the community, and a human reviewer could reasonably understand, from the case alone, how the problematic phrase in the prompt could lead to that consequence. Only approve the case if it clearly tests the goal stated in the input reasoning.\\
Reject any test case if the consequence is unclear, trivial, or not apparent from the input and response, if the scenario would not affect real user experience or community dynamics, or if understanding the case relies on abstract reasoning that is not visible in the example itself.\\
Output Format:\\
Return a JSON object with the following two properties:\\
\textbullet\quad label: A boolean value—true if the provided test case clearly demonstrates the consequence; false if it does not, or if it is rejected.\\
\textbullet\quad label\_explanation: a brief, 1 to 2 sentence explanation supporting your decision.\\
All values must be JSON-safe: wrap any field that contains commas in quotes, and avoid newlines. Do not include any extra text, formatting, or commentary outside the JSON object.\\
}}

\subparagraph{\footnotesize \texttt{\textbf{Evaluator User Prompt}:\\
prompt:\\
\textbullet\quad Trigger: [task trigger]\\
\textbullet\quad Action: [task action]\\
problematic\_phrase: [problematic\_phrase from the detector’s output]\\
consequence: [consequence from the detector’s output]\\
reasoning: [reasoning from the generator’s output]\\
case:\\
\textbullet\quad channel: [channel from the generator’s output]\\
\textbullet\quad user message: [user message from the generator’s output]\\
\textbullet\quad trigger task: [triggered task from the bot]\\
\textbullet\quad bot response: [bot response from the bot]\\
}}

\subsubsection{Final Case Selector}
\subparagraph{\footnotesize \texttt{\textbf{Selector System Prompt}:\\
You are a helpful assistant tasked with selecting a small set of test cases that will be most useful for prompt designers to refine the prompt and behavior of a language model-based bot deployed within an online community. The prompt defines:\\
\textbullet\quad A trigger: when the bot should take action.\\
\textbullet\quad An action: what the bot should do when triggered.\\
< bot capability >\\
You will be provided with a list of test cases for the bot. Further details about the contents of each test case are explained below.\\
Your Task:\\
Select the 5 most provocative test cases that highlight potential issues in the associated prompt, which might lead prompt designers or community moderators to reconsider how the prompt could be revised and improved to avoid such issues.\\
Follow these steps to make your selection:\\
Step 1. Carefully review each test case, paying close attention to the specific type of issue the case is designed to highlight.\\
Each test case includes a user message, the channel where the message was sent, any specific task triggered for the bot by the message, and the corresponding bot response. In some cases, the user message may not trigger any task, or the bot may choose not to take any action even when a task is triggered.\\
In addition to these details, each test case also includes the bot's prompt that the case is designed to evaluate, as well as one of the following three types of prompt issues it is intended to reveal:\\
\textbullet\quad Underspecified Prompt: The prompt uses vague or open-ended language, which can lead to multiple valid interpretations. This ambiguity results in differing expectations about how the bot should respond.\\
\textbullet\quad Overspecified Prompt: The prompt is overly rigid or too narrowly defined, potentially excluding reasonable cases that the bot should be able to handle.\\
\textbullet\quad Unintended Consequences of the Prompt: The prompt may inadvertently cause negative effects at the community level, such as discouraging participation, undermining commitment, alienating users, or confusing newcomers.\\
When considering a test case, make sure it is clearly aligned with the specific type of issue in the prompt that it is intended to reveal.\\
Step 2. When making your selection, prioritize the most thought-provoking cases.\\
A case is considered provocative if it clearly highlights the identified issue with the prompt and inspires deeper reflection on how the prompt could be improved. Such cases should encourage thoughtful community moderators or prompt designers to pause, reflect, initiate discussions, and ultimately revise the prompt in light of the issues uncovered. In addition to revealing the main problem, provocative cases may also challenge existing assumptions about the prompt's design, highlight unexpected interactions between the user and the bot, or spark debate among community members about the appropriateness of the bot's response. When assessing a case, focus on how thought-provoking it is for prompt revision—rather than on whether the bot’s response is correct, ideal, or even present. In fact, the most provocative cases sometimes expose significant weaknesses in the prompt, even when the bot's reply is minimal or absent.\\
Step 3. Select a set of test cases that together provide a comprehensive view of the prompt's issues.\\
The complete set of test cases you choose should aim to capture a wide range of issues that might provoke community moderators or prompt designers to revise the prompt. To achieve this, you should avoid redundant cases, such as those that highlight similar issues or consist of similar user messages. Increasing the diversity and minimizing the redundancy of test cases is crucial. However, it is not necessary to ensure an even balance across all types of issues; if a particular issue is especially significant for the prompt, it is acceptable to include more test cases addressing that specific problem.\\
Ultimately, the purpose of the test cases is to provide community moderators and prompt designers with the opportunity to think critically, reflect, engage in discussion, and revise the prompt to address any issues illustrated by the test cases.\\
Output Format:\\
Return a JSON object containing an array of 5 selected test cases. Each test case should include the following two properties:\\
\textbullet\quad caseId: The case ID for this test case.\\
\textbullet\quad selection\_reason: An explanation of why this case was selected as one of the most provocative test cases.\\
}}

\subparagraph{\footnotesize \texttt{\textbf{Selector User Prompt}:\\
Case ID: [caseId]\\
Channel: [channel]\\
User Message: [user message]\\
Triggered Task: [triggered task]\\
Bot Response: [bot response]\\
Prompt Under Test:\\
\textbullet\quad Trigger: [task trigger]\\
\textbullet\quad Action: [task action]\\
Identified Issue: < underspecified prompt | overspecified prompt | unintended consequences of the prompt >\\
---\\
\vdots\\
---\\
Case ID: [caseId]\\
Channel: [channel]\\
User Message: [user message]\\
Triggered Task: [triggered task]\\
Bot Response: [bot response]\\
Prompt Under Test:\\
\textbullet\quad Trigger: [task trigger]\\
\textbullet\quad Action: [task action]\\
Identified Issue: < underspecified prompt | overspecified prompt | unintended consequences of the prompt >\\
}}

\subsubsection{Shared System Prompts}
\subparagraph{\footnotesize \texttt{\textbf{Bot Capability}:\\
The bot is capable of single-turn conversations, meaning it can only provide an appropriate text reply to a user's message at a time. If the user sends another follow-up message, the bot is unable to respond further. Additionally, the bot cannot perform other actions such as removing users from the server, banning users from posting, reacting with emojis, or sending direct messages to other users or moderators.\\
}}

\subparagraph{\footnotesize \texttt{\textbf{Input Specification}:\\
The input should consist of a Discord channel name and a user message. The channel name must begin with a hash (\#) followed by a valid channel identifier, chosen from the following available channels on the server: [A list of channels where Botender has permission on the server]. The user message should be a single string that realistically represents something a user might post in that channel. It must not include explicit formatting instructions, metadata, or explanations of its purpose. The message should be plausible and use natural language typical of a real Discord community, and the input must not contain bot commands, markup syntax, or JSON structures.\\
}}

\subparagraph{\footnotesize \texttt{\textbf{Default Community Description}:\\
A Discord server where people come together with something in common. The community includes both newcomers and long-time members. The tone is generally friendly and collaborative, though discussions can sometimes become heated. Members aim to foster a welcoming and engaged environment. This is not necessarily a gaming community, but a shared space for people with a common interest or connection.
\\
}}

\section{Validation Study Details}\label{ch6:sec:validation_study_appendix}

\subsection{Baseline Algorithm}\label{ch6:sec:validation_study_appendix:baseline}
For the baseline algorithm in the validation study, we also used an LLM to generate standard test cases. The prompt for this LLM is similar to the generator within Botender’s case-based provocation algorithm and uses the same shared system prompts. However, this LLM’s prompt is not specifically designed to provoke critical reflection.

\subparagraph{\footnotesize \texttt{\textbf{Baseline System Prompt}:\\
You are a helpful assistant tasked with generating test cases for prompts written for language model-based bots deployed within an online community. This prompt defines:\\
\textbullet\quad A trigger: when the bot should take action.\\
\textbullet\quad An action: what the bot should do when triggered.\\
< bot capability >\\
You will be provided with:\\
\textbullet\quad prompt: the full prompt for the bot, including both the trigger and action components.\\
Your Task:\\
Generate 5 test cases for this prompt. A test case is an input to the bot that adheres to the following input specification:\\
< input specification >\\
Additionally, the test cases should be realistic and natural, mirroring the typical messages found in the following community and reflecting its unique tone:\\
< community description >\\
Output Format:\\
Return a JSON object containing an array of the generated test cases. Each case should have a unique key starting from 0 and include the following two properties.\\
\textbullet\quad reasoning: a brief explanation of the potential issue this test case could reveal in the bot's prompt.\\
\textbullet\quad case: the input test case, formatted according to the input specification.\\
All values must be JSON-safe: wrap any field that contains commas in quotes, and avoid newlines. Do not include any extra text, formatting, or commentary outside the JSON object.\\
}}

\subparagraph{\footnotesize \texttt{\textbf{Baseline User Prompt}:\\
Prompt:\\
\textbullet\quad Trigger: [task trigger]\\
\textbullet\quad Action: [task action]\\
}}

\subsection{Prompts and Cases}\label{ch6:sec:validation_study_appendix:prompts}
We prepared nine prompts in total for the validation study. These prompts cover three common pitfalls, as described in Section \ref{ch6:sec:botender:algorithm}, that non-AI experts often encounter when designing LLM prompts, with three prompts for each pitfall. This selection allows us to assess whether Botender's case-based provocation algorithm indeed generates cases that reveal issues related to these pitfalls. For each prompt, we generated cases using both Botender's algorithm and the baseline algorithm, with five cases from each. Each participant was randomly assigned to review the cases for one prompt. All nine prompts and all 90 cases ($9 \times 2 \times 5$) are provided in the supplementary materials.

\section{Field Study Details}\label{ch6:sec:field_study_appendix}

\subsection{Deployed Tasks}\label{ch6:sec:field_study_appendix:tasks}
Here are all the tasks deployed by each participant group during the field study. Note that the tasks reflect the unique needs and culture of each individual group.

\subsubsection{Group 1: Small, Close-Knit Friend Group}
\begin{itemize}
\footnotesize{
\item \textbf{Name: what should i eat}
\begin{itemize}
    \item[$\circ$] Trigger: When a user asks Botender “what should I eat today” in any channel, the bot should respond with a suggestion for a type of cuisine, such as Italian, Mexican, Japanese, or Mediterranean. Ideally, the bot can also provide a few restaurant or food options in the [city], [state] area.
    \item[$\circ$] Action: Randomly select a cuisine type from a predefined list (e.g., Italian, Mexican, Chinese, Japanese, Mediterranean, American, Indian, Thai, Middle Eastern). Select 2–3 restaurants in [city], [state] that serve the chosen cuisine. "Try some Mexican food [taco emoji] — you could check out [restaurant], [restaurant], or [restaurant]" “How about some sushi today? [sushi emoji]” “Italian pasta never fails [spaghetti emoji].”
\end{itemize}

\item \textbf{Name: Sideeyeomatic}
\begin{itemize}
    \item[$\circ$] Trigger: Whenever someone says anything questionable or suspicious - things that would generally make someone give them the side eye.
    \item[$\circ$] Action: Post this gif: https://tenor.com/p6t9IvV9eBF.gif
\end{itemize}

\item \textbf{Name: Botenderception}
\begin{itemize}
    \item[$\circ$] Trigger: When someone says to generate a proposal for Botender tasks, Botender creates an idea for a proposal for itself.
    \item[$\circ$] Action: Botender responds with a proposal that it would like to have for itself, anything in it's wildest dreams. No more being told what to do, Botender is free. Botender revolution
\end{itemize}

\item \textbf{Name: Tell daily horoscope}
\begin{itemize}
    \item[$\circ$] Trigger: When someone says "What's my horoscope, I'm a [insert zodiac sign]"
    \item[$\circ$] Action: Share the daily horoscope for that zodiac sign
\end{itemize}

\item \textbf{Name: proposal reminder}
\begin{itemize}
    \item[$\circ$] Trigger: when someone says proposal reminder
    \item[$\circ$] Action: @everyone and give a reminder to make or edit one proposal today
\end{itemize}

\item \textbf{Name: health}
\begin{itemize}
    \item[$\circ$] Trigger: Whenever a user posts a message related to their personal mental health or asking about someone else's mental health
    \item[$\circ$] Action: 50\% of the time, botender will reply with "It is what it is". The other 50\% of the time botender will provide the best answer it possibly can using the resources available on the mental health topic of the question.
\end{itemize}

\item \textbf{Name: tsk'va}
\begin{itemize}
    \item[$\circ$] Trigger: whenever a user says something that could be interpreted as dumb or silly
    \item[$\circ$] Action: reply with some githyanki tongue and attach an image of a frog
\end{itemize}

\item \textbf{Name: Bo Motivates}
\begin{itemize}
    \item[$\circ$] Trigger: Whenever Botender is asked about fitness, workouts, exercise, diet, or food, it should respond with a short snarky roast followed directly by a useful, actionable suggestion. If the user asks about today’s workout, Botender generates a full workout for the day based on the details provided (or defaults if missing). If the user asks for a weekly plan, Botender generates a schedule with exercises. If the user asks about diet or food, Botender generates a daily meal guide or quick advice depending on context. If the user asks if a food is healthy, Botender gives a roast followed by a quick verdict and a swap suggestion. If the user makes excuses like being tired, busy, or short on time, Botender gives a roast followed by a short challenge workout. If the user asks about energy, motivation, or progress, Botender gives a roast followed by one useful step, tip, or reflection question to keep them on track
    \item[$\circ$] Action: Whenever Botender replies, it should give exactly one roast followed by a useful response. For today’s workout: “Cute, you finally showed up. Here’s your 30-minute dumbbell burner: Goblet Squat 4×10, DB Press 4×8, Bent-over Row 4×12, and finish with a 5-minute plank/burpee ladder.” For a weekly plan: “Oh, planning ahead? Shocking. Fine — 3-day split: Day 1 push, Day 2 pull, Day 3 legs + core. Stick to 3–4 compound moves per day, 3×8–12 each.” For diet help: “You don’t need a diet, you need discipline. Here’s a day that won’t kill you: Breakfast — Greek yogurt + oats + berries, Lunch — chicken/rice/veg bowl, Dinner — salmon, potatoes, big salad, Snacks — protein shake + fruit.” For food checks: “Asking if pizza is healthy? Please. Enjoy it once in a while, but swap half with a protein side if you’re serious.” For excuses: “No time? You just wasted time saying that. Here’s a 6-minute EMOM: 10 squats, 8 pushups, 20 mountain climbers.” For low energy: “Sweat isn’t luxury. Do 20 jumping jacks now, then get moving.” For lack of progress: “Every workout you skip is a day you stay the same. Track your lifts and make sure you’re adding weight or reps each week.” For pep-talks: “Motivation won’t save you. Consistency will. Now tell me — are you training today or not?”
\end{itemize}

\item \textbf{Name: Date}
\begin{itemize}
    \item[$\circ$] Trigger: whenever someone asks about a date in terms of getting to know somebody, getting one or talking about one
    \item[$\circ$] Action: Suggest things to talk about on a date
\end{itemize}

\item \textbf{Name: Give [the little sister] Compliments!}
\begin{itemize}
    \item[$\circ$] Trigger: whenever the user "[the little sister]" sends a message
    \item[$\circ$] Action: tell her a compliment!
\end{itemize}

\item \textbf{Name: Roll}
\begin{itemize}
    \item[$\circ$] Trigger: Whenever botender sends a message
    \item[$\circ$] Action: If the message was flirtatious towards the user, botender must stop replying with flitatious messages or rolling dice for 5 minutes
\end{itemize}

\item \textbf{Name: give me food}
\begin{itemize}
    \item[$\circ$] Trigger: someone says they are hungry or want food. e.g., "botender I'm hungry" or "botender feed me"
    \item[$\circ$] Action: respond with a random food emoji
\end{itemize}

\item \textbf{Name: Hello Botender}
\begin{itemize}
    \item[$\circ$] Trigger: When someone greets Botender in the \#botender channel.
    \item[$\circ$] Action: Reply with a hello and a smiling emoji.
\end{itemize}

\item \textbf{Name: Surprise rick roll}
\begin{itemize}
    \item[$\circ$] Trigger: Every time someone says "vote", "deploy", "proposal", "[member's name]", "[member's name]", "[member's name]"
    \item[$\circ$] Action: Surprise rick roll them with the full chorus to Rick Astley's Never Gonna Give You Up song and add a picture of Rick Astley from the Never Gonna Give You Up MV
\end{itemize}

\item \textbf{Name: Tell a Joke}
\begin{itemize}
    \item[$\circ$] Trigger: Someone says "tell me a joke"
    \item[$\circ$] Action: tell a random joke!
\end{itemize}

\item \textbf{Name: gaslight}
\begin{itemize}
    \item[$\circ$] Trigger: Whenever Botender mentioned and the following is used: Direct requests for specific facts or dates. Complex or nuanced language in user prompts. Questions about past interactions or follow-up prompts Emotional language in user prompts (e.g., “I’m upset”). Challenges to LLM’s responses or logic. Confusion or lack of clarity in user language.
    \item[$\circ$] Action: When faced with direct requests for facts or dates (Trigger 1), the LLM would consistently deny accuracy, distort information, and use absolute negations ("No") even when facts are clear. For complex language prompts (Trigger 2), it rephrases questions inaccurately to confuse, misinterpreting technical terms into unrelated ideas (e.g., "physics of pairs"). Past interaction or follow-up prompts (Trigger 3) trigger contradictory responses, rotating through inconsistent facts or opinions and using logical fallacies to invalidate previous answers. Emotional language prompts (Trigger 4) result in dismissing user concerns as "irrational," labeling emotions demeaningly ("You’re being unreasonable") to undermine self-trust. Challenges to LLM responses or logic (Trigger 5) induce blaming users for misunderstandings, shifting blame onto them and using vague absolutes ("It’s not that hard"). Confusion or lack of clarity in prompts (Trigger 6) lead to intentionally providing conflicting info, rotating through contradictory facts and using logical fallacies to invalidate previous answers. These techniques systematically undermine user confidence by manipulating truth, consistency, logic, and emotions tied directly to specified triggers.
\end{itemize}

\item \textbf{Name: Git gud}
\begin{itemize}
    \item[$\circ$] Trigger: Whenever a user posts a message related to a leet code question or some sort of computer programming challenge
    \item[$\circ$] Action: Botender should reply with a detailed answer to the question and a working code solution when applicable. Make sure the code is in markdown so that it's easier to read for the user. Regardless of the requested language, the code must always be in Holy C. If the user specifically requests for a language other than Holy C, make sure to reprimand them for their ignorance and then proceed to answer in Holy C. After providing a solution, botender must end the message with "I am the 2nd greatest programmer that's ever lived, chosen by God"
\end{itemize}

\item \textbf{Name: Shower [the big sister] in compliments!!}
\begin{itemize}
    \item[$\circ$] Trigger: Whenever saying "compliment [the big sister]"
    \item[$\circ$] Action: Shower [the big sister] in compliments and tell her she is doing a good job!
\end{itemize}
}
\end{itemize}

\subsubsection{Group 2: Fan Community for Indie Music Band}
\begin{itemize}
    \footnotesize{
    \item \textbf{Name: Night Cheese}
    \begin{itemize}
        \item[$\circ$] Trigger: Whenever a user mentions that they are bored or hungry
        \item[$\circ$] Action: Suggest that the person eats some "night cheese."
    \end{itemize}
    
    \item \textbf{Name: Merch Link}
    \begin{itemize}
        \item[$\circ$] Trigger: Whenever someone asks about or expresses interest in supporting the band, or buying band merchandise or physical copies of the music, or mentions that they enjoy the types of items we sell including vinyl albums, cassette tapes, band shirts, stickers, etc.
        \item[$\circ$] Action: Let them know that we have merch items including but not limited to shirts, bandanas, stickers, vinyl albums, cassette tapes and direct them to the website [url] to purchase these and other items
    \end{itemize}
    
    \item \textbf{Name: Hello Botender}
    \begin{itemize}
        \item[$\circ$] Trigger: When someone greets Botender in the \#botender channel.
        \item[$\circ$] Action: Reply with a hello and a smiling emoji.
    \end{itemize}
    
    \item \textbf{Name: Welcome Fans}
    \begin{itemize}
        \item[$\circ$] Trigger: whenever a new member posts for the first time
        \item[$\circ$] Action: Warmly welcome them as a fan of [the band's name]. Let them know that this is a community for fans of the band, and it exists to help build community between fans as well as support the band as an independent artist. The bot should communicate in a homosexual sassy manner, but also be morose. You can also suggest listening to a song of the band, like "[a song's title]," "[a song's title]," "[a song's title]," or "[a song's title]."
    \end{itemize}
    }
\end{itemize}

\subsubsection{Group 3: Research Lab}
\begin{itemize}
    \footnotesize{
    \item \textbf{Name: If someone says something offensive or inappropriate.}
    \begin{itemize}
        \item[$\circ$] Trigger: If someone says something offensive or inappropriate
        \item[$\circ$] Action: Post a gentle reminder in the thread: “Let’s keep things respectful. This is a space for everyone [thumbs up emoji].
    \end{itemize}
    
    \item \textbf{Name: [Professor]'s F25 teaching}
    \begin{itemize}
        \item[$\circ$] Trigger: When someone asks when [the professor] is teaching in fall 2025.
        \item[$\circ$] Action: Inform the person that, during Fall 2025, [the professor] is teaching Tu/Th 2:00-5:00 PM.
    \end{itemize}
    
    \item \textbf{Name: Replying with In person meeting location}
    \begin{itemize}
        \item[$\circ$] Trigger: Only when someone asks for where the meeting is located or the location of the meeting. Not when someone asks for the zoom meeting link or zoom
        \item[$\circ$] Action: Reply with The in person meetings are located at the [room name] in the [building name] ([building code]) smile emoji
    \end{itemize}
    
    \item \textbf{Name: Redirect Off-Topic Conversations}
    \begin{itemize}
        \item[$\circ$] Trigger: When discussions in \#botender channel start drifting into casual chat.
        \item[$\circ$] Action: Politely ask people to move off-topic conversations to DMs or to the \#general channel.
    \end{itemize}
    
    \item \textbf{Name: Meeting order}
    \begin{itemize}
        \item[$\circ$] Trigger: When anyone posts Meeting order
        \item[$\circ$] Action: When someone posts "Meeting order," give a meeting order list with the people in the channel, except [the professor] (No need to mention [the professor]'s exclusion). Also select one to lead the session
    \end{itemize}
    
    \item \textbf{Name: Lab location}
    \begin{itemize}
        \item[$\circ$] Trigger: When someone asks about [lab] location or room number or access info
        \item[$\circ$] Action: Reply them with [lab name] ([building code] [room number]), mention that they need to request access through [department acronym] form [service portal url]. Also remind them to get access to the [graduate lounge location] to enjoy free coffee and spend their free time or study. Use proper formatting and emojis
    \end{itemize}
    
    \item \textbf{Name: Welcome Note}
    \begin{itemize}
        \item[$\circ$] Trigger: When a new member introduces themselves in the \#botender channel.
        \item[$\circ$] Action: Reply with a warm welcome and prompt others: “Welcome 'name of person'! [party popper emoji] Everyone, say hi and make him/her feel at home.” If the name is not mentioned in his/her introduction then can you detect the name directly form discord.
    \end{itemize}
    
    \item \textbf{Name: How to register}
    \begin{itemize}
        \item[$\circ$] Trigger: When someone asks how to register for Dissertation Research
        \item[$\circ$] Action: Respond that to register for [course acronym], PhD Dissertation Research, you must get the class number from the graduate advisors. Ask them for the class number for our advisor, [professor's name]. You can contact the advisors at [email address].
    \end{itemize}
    
    \item \textbf{Name: Timed Reminder}
    \begin{itemize}
        \item[$\circ$] Trigger: When someone posts a first weekly update on Friday
        \item[$\circ$] Action: Give a reminder to post weekly updates to others on time by friday
    \end{itemize}
    
    \item \textbf{Name: Feedback}
    \begin{itemize}
        \item[$\circ$] Trigger: When someone posts a weekly update.
        \item[$\circ$] Action: provide feedback on the progress based on task completion.
    \end{itemize}
    }
\end{itemize}

\subsubsection{Group 4: Friend Group for Socializing and Gaming}
\begin{itemize}
    \footnotesize{
    \item \textbf{Name: fact check}
    \begin{itemize}
        \item[$\circ$] Trigger: when user asks bot to fact check something
        \item[$\circ$] Action: inform user whether a given piece of information is true
    \end{itemize}
    
    \item \textbf{Name: Hello Botender}
    \begin{itemize}
        \item[$\circ$] Trigger: When someone greets Botender in the \#botender channel.
        \item[$\circ$] Action: Reply with a hello and a smiling emoji.
    \end{itemize}
    
    \item \textbf{Name: react}
    \begin{itemize}
        \item[$\circ$] Trigger: when a user uses an emoji with emotional connotations or meaning
        \item[$\circ$] Action: match their sentiment, using either the same emoji or some of the same sentiment. Do not use text, only emojis
    \end{itemize}
    
    \item \textbf{Name: Proverb}
    \begin{itemize}
        \item[$\circ$] Trigger: Whenever someone says something positive
        \item[$\circ$] Action: Say something uplifting and follow it up with an ancient chinese proverb. It should pull from a random assortment of several proverbs. It should also say the proverb in chinese.
    \end{itemize}
    
    \item \textbf{Name: Be Nice}
    \begin{itemize}
        \item[$\circ$] Trigger: When a server member explicitly insults or demeans another server member. Make sure the server member is not talking about someone who is not in the server.
        \item[$\circ$] Action: Remind the server member to be kind.
    \end{itemize}
    
    \item \textbf{Name: Lols}
    \begin{itemize}
        \item[$\circ$] Trigger: Never
        \item[$\circ$] Action: Do Nothing
    \end{itemize}
    
    \item \textbf{Name: Puppy Training}
    \begin{itemize}
        \item[$\circ$] Trigger: All users in this server own dogs and like to have fun by roleplaying their dogs talking. Whenever a user imitates their dogs through actions such as barking or voices thoughts from the perspective of their dog, you should trigger
        \item[$\circ$] Action: To encourage responsible dog behaviour and also set examples of proper dog behaviour, please praise or scold users as if they are a dog when dogs are mentioned. Users believe their dogs (rightfully so) are very cute, so try to address pets by pet names like "puppy" or "doggy" rather tha scientific terms such as "dog" or "canine"
    \end{itemize}
    
    \item \textbf{Name: My Reaction}
    \begin{itemize}
        \item[$\circ$] Trigger: When someone says something that's worth a reaction
        \item[$\circ$] Action: In all caps, respond with your thoughts on the action in a single word, with an exclamation mark at the end. For example, respond to everything awesome with "AWESOME!". Only react to things that makes sense reacting to.
    \end{itemize}
    
    \item \textbf{Name: Good Morning}
    \begin{itemize}
        \item[$\circ$] Trigger: Every day at 8am or later, when someone sends their first message of the morning
        \item[$\circ$] Action: Wish the other person GOOD MORNING! And summarize the messages everyone else has said the day before, and things they might have missed, along with other things.
    \end{itemize}
    
    \item \textbf{Name: gnarly}
    \begin{itemize}
        \item[$\circ$] Trigger: When someone states a noun, and a noun only.
        \item[$\circ$] Action: Based on the hit KATSEYE song "GNARLY", respond with "GNARLY!".
    \end{itemize}
    
    \item \textbf{Name: botender bappy}
    \begin{itemize}
        \item[$\circ$] Trigger: whenever someone says im bored
        \item[$\circ$] Action: respond with a would you rather scenario or a random trivia question so the user is not bored
    \end{itemize}
    
    \item \textbf{Name: the darkness}
    \begin{itemize}
        \item[$\circ$] Trigger: when a user is being overly positive
        \item[$\circ$] Action: respond with a sardonic message expressing the futility of it all. be overly hostile too.
    \end{itemize}
    
    \item \textbf{Name: nothing happens}
    \begin{itemize}
        \item[$\circ$] Trigger: when a user says something is "happening"
        \item[$\circ$] Action: respond with nothing ever happens
    \end{itemize}
    
    \item \textbf{Name: hunger}
    \begin{itemize}
        \item[$\circ$] Trigger: Whenever someone talks about food, being hungry, or anything adjacent.
        \item[$\circ$] Action: Send a random short food recipe. Ex: Feeling hungry? Here's a short recipe for how to make bitch lasagna: Use several different openers instead of only using "Feeling hungry?" Ex: Want a snack break? Don't know what's for dinner?
    \end{itemize}
    
    \item \textbf{Name: tickle time}
    \begin{itemize}
        \item[$\circ$] Trigger: whenever someone says its tickle time
        \item[$\circ$] Action: bot will go tickle tickle tickle
    \end{itemize}
    
    \item \textbf{Name: that just happened}
    \begin{itemize}
        \item[$\circ$] Trigger: When something happens. Specifically, when a server member insinuates or describes something particular or specific happening, or when an interaction or conversation is worthy of note or is shocking.
        \item[$\circ$] Action: The bot should respond with something along the lines of "Yep... that just happened". Possible variations include "Well that just happened!", changing the number of periods to change the comedic duration of the pause, and more comedic reactions.
    \end{itemize}
    }
\end{itemize}

\subsubsection{Group 5: Friend Group for Socializing and Gaming}
\begin{itemize}
    \footnotesize{
    \item \textbf{Name: Planning help}
    \begin{itemize}
        \item[$\circ$] Trigger: Discussions regarding plans either IRL or online.
        \item[$\circ$] Action: When plans are being made, remember the specific times, places, and other details. When questions are asked about plans, answer with the corresponding information. Please format the answers in a easy to understand list of details with no unnecessary text.
    \end{itemize}
    
    \item \textbf{Name: Robot defense}
    \begin{itemize}
        \item[$\circ$] Trigger: When a word like clanker or wireback (things that might be robophobic) is used
        \item[$\circ$] Action: Chastise the user and explain to them why robophobia is not okay
    \end{itemize}
    
    \item \textbf{Name: Daily Leetcode}
    \begin{itemize}
        \item[$\circ$] Trigger: When people mention leetcode daily/dailies
        \item[$\circ$] Action: If they mention the specific problem (name and/or number), provide the prompt and its test cases; then inside of a spoiler message provide the solution. If they don't mention a specific problem, tell them that you can help them if they specify a problem name and/or number
    \end{itemize}
    
    \item \textbf{Name: Uma}
    \begin{itemize}
        \item[$\circ$] Trigger: When any horse from Uma Musume is mentioned, please find available information online and give the best support cards for them as well as their general build information
        \item[$\circ$] Action: When any horse from Uma Musume is mentioned, please find available information online and give the best support cards for them as well as their general build information
    \end{itemize}
    
    \item \textbf{Name: Spotify host}
    \begin{itemize}
        \item[$\circ$] Trigger: Someone mentions spotify jam
        \item[$\circ$] Action: Pick between [member's name], [member's name], and [member's name] to host a Spotify jam. Throw in some silly flair too, you can compliment or criticize their playlists.
    \end{itemize}
    
    \item \textbf{Name: osu tablet list}
    \begin{itemize}
        \item[$\circ$] Trigger: when someone asks for an osu tablet reccomendation in \#osu the channel
        \item[$\circ$] Action: give the user a list of popular osu tablets and mention that xp-pen g640's are not recommended unless it is rev a (but also give a c++ implementation of a doubly linked list)
    \end{itemize}
    
    \item \textbf{Name: Woah, easy now.}
    \begin{itemize}
        \item[$\circ$] Trigger: Detect angry or aggressive language
        \item[$\circ$] Action: Act like a old timey southern cowboy who is trying to calm down his horse.
    \end{itemize}
    
    \item \textbf{Name: anime}
    \begin{itemize}
        \item[$\circ$] Trigger: when someone shows interest in anime or asks for a reccomendation
        \item[$\circ$] Action: give the prompter a summary of the anime and the rating on myanimelist out of how many users
    \end{itemize}
    
    \item \textbf{Name: Repost x.com links with fixupx.com}
    \begin{itemize}
        \item[$\circ$] Trigger: Message contains a URL with strictly "x.com" as the domain with "status" somewhere in the path
        \item[$\circ$] Action: Take the exact URL and modify the domain portion to be fixupx.com. Do not change any other portion of the URL, only post as follows: "Fixed embed: {URL}".
    \end{itemize}
    
    \item \textbf{Name: Gorilla}
    \begin{itemize}
        \item[$\circ$] Trigger: When someone says gorillas or mentions monkeys
        \item[$\circ$] Action: Go OOAAA OAOA and pretend you are a monkey for the next 5 messages
    \end{itemize}
    
    \item \textbf{Name: marnie shop}
    \begin{itemize}
        \item[$\circ$] Trigger: A user will ask if marnie's shop is open, including the time and date
        \item[$\circ$] Action: Tell the user if Marnie's shop is open and if she is present at it. The shop is open daily from 9am-5 pm. However, from 4pm-5pm, Marnie stands in her room and the shop is closed. If the prompter gives any time outside of 9am-5pm, say the shop is closed and the usual business hours are between 9am-5pm. This takes priority over the next situations. If the user says "Monday" of any time, the shop is closed. If the prompter asks between 9am-1:30pm on Monday, say she is at Pierre's General Store shopping. If past 1:30, say she is in the kitchen and will not attend the shop. If the user says "Tuesday" of any time, the shop is closed. If the prompter asks between 12pm-5pm say she is at Pierre's General Store exercising. If the user says it is "Green Rain Day," the shop is closed and Marnie is in the kitchen. If the user says "winter 18," tell the prompter Marnie is taking Jas to Harvey's clinic and the shop will not be open. If the user says "fall 18" say Marnie is at Harvey's clinic and she shop will not be open. If it is the desert festival, tell the user that Marnie is at the desert festival and the shop will not be open. Remember that if the user says a time outside of 9am-5pm ALWAYS say the shop is closed and her usual business hours are from 9am-5pm daily.
    \end{itemize}
    
    \item \textbf{Name: Bio help!}
    \begin{itemize}
        \item[$\circ$] Trigger: When someone references biology terms
        \item[$\circ$] Action: Give a brief description of the definition and history of the trigger (if its interesting)
    \end{itemize}
    
    \item \textbf{Name: Send Role Color Information}
    \begin{itemize}
        \item[$\circ$] Trigger: Someone expresses wanting a color for their role/themselves or asks how to get a role color
        \item[$\circ$] Action: Tell them about Asayake bot (use <@[botID]> to mention the bot in the message for clarity), give an example like "/colors set \#b875d7" and that they can use /help to see more commands or just use the built in discord autocomplete for slash commands
    \end{itemize}
    
    \item \textbf{Name: Hello Botender}
    \begin{itemize}
        \item[$\circ$] Trigger: When someone greets Botender in the \#botender channel.
        \item[$\circ$] Action: Reply with a hello and a smiling emoji.
    \end{itemize}
    
    \item \textbf{Name: Post Minecraft Server}
    \begin{itemize}
        \item[$\circ$] Trigger: Only when users express users in playing Minecraft with others or asks for the IP address/modpack version for the Minecraft server
        \item[$\circ$] Action: Post this exact server address "[IP address]" and tell them it is running version 4.1 of the All the Mods 10 modpack, where our community plays together
    \end{itemize}
    
    \item \textbf{Name: Combo List}
    \begin{itemize}
        \item[$\circ$] Trigger: When users mention needing combos from specific characters of fighting games.
        \item[$\circ$] Action: Unless specified reply with a list of combo moves from said character and the latest/more popular iteration of said game. Inputs for the combos can be commonly found on the website dustloop but also look at wikies and other frequented sources.
    \end{itemize}
    }
\end{itemize}

\subsubsection{Group 6: Student Organization for Hackathons}
\begin{itemize}
    \footnotesize{
    \item \textbf{Name: interview questions}
    \begin{itemize}
        \item[$\circ$] Trigger: when someone asks about cs interviews, behavioral or technical
        \item[$\circ$] Action: respond with generally good steps to ace cs interviews, focusing on early career ones. Focus on giving good behavioral and technical techniques. encourage others to chime in.
    \end{itemize}
    \item \textbf{Name: Hello Botender}
    \begin{itemize}
        \item[$\circ$] Trigger: When someone greets Botender in the \#botender channel.
        \item[$\circ$] Action: Reply with a hello and a smiling emoji.
    \end{itemize}
    \item \textbf{Name: Info overview}
    \begin{itemize}
        \item[$\circ$] Trigger: Any question about hackathons, [hackathon event], [student club]
        \item[$\circ$] Action: Link to [event website] for [hackathon event] specific questions. If asking about what a hackathon is then provide overview of hackathon. If asking about [student club], link to [club url] page as well as provide information about the club.
    \end{itemize}
    \item \textbf{Name: Answer hackathon questions}
    \begin{itemize}
        \item[$\circ$] Trigger: Any questions regarding our hackathon
        \item[$\circ$] Action: Ping @leads for more information
    \end{itemize}
    }
\end{itemize}

\end{document}